\documentclass[acmsmall]{acmart}
\AtBeginDocument{%
  }

\usepackage{algorithmic}
\usepackage{graphicx}
\usepackage{xcolor}
\usepackage{soul}
\usepackage{colortbl}
    
\usepackage{multirow}
\usepackage{multicol}
\usepackage[normalem]{ulem}
\usepackage{ragged2e}
\usepackage{subcaption}
\usepackage{listings}

\sethlcolor{yellow}
\newcommand{\new}[1]{#1}

\newcommand{\old}[1]{}

\newcommand{\cedrdata}{\emph{hete\_Data}}
\newcommand{\cedrmalloc}{\emph{hete\_Malloc}}
\newcommand{\cedrfree}{\emph{hete\_Free}}
\newcommand{\cedrsync}{\emph{hete\_Sync}}

\definecolor{dkgreen}{rgb}{0,0.6,0}
\definecolor{lgrey}{rgb}{0.9,0.9,0.9}
\lstdefinelanguage{cpp}{
      backgroundcolor=\color{lgrey},  
      basicstyle=\footnotesize \ttfamily \color{black} \bfseries,   
      breakatwhitespace=false,       
      breaklines=true,               
      captionpos=b,                   
      commentstyle=\color{dkgreen},   
      deletekeywords={...},          
      escapeinside={\%*}{*)},                  
      frame=single,                  
      language=C++,                
      keywordstyle=\color{purple},  
      morekeywords={size_t,FFT,allocate,deallocate,complex,uint32_t,CEDR_FFT,malloc,free,fft_wrapper,cufftComplex,cudaMalloc,cudaMemcpy,cudaDeviceSynchronize,cudaFree,cuda_fft_wrapper,iris_mem_release,iris_synchronize,iris_task_submit,iris_task_d2h,iris_task_h2d,iris_mem_create,iris_task,iris_mem,printResults}, 
      identifierstyle=\color{black},
      stringstyle=\color{blue},      
      numbers=left,                 
      numbersep=5pt,                  
      numberstyle=\tiny\color{black}, 
      rulecolor=\color{black},        
      showspaces=false,               
      showstringspaces=false,        
      showtabs=false,                
      stepnumber=1,                   
      tabsize=5,                     
      title=\lstname,
      moredelim=**[is][\color{red}\bfseries]{\@}{\@},
      moredelim=**[is][\color{blue}\bfseries]{\$}{\$},
    }

\begin{document}

\title{RIMMS: Runtime Integrated Memory Management System for Heterogeneous Computing}

\author{Serhan Gener}
\affiliation{%
  \institution{University of Arizona}
  \city{Tucson}
  \state{Arizona}
  \country{USA}
}
\email{gener@arizona.edu}
\orcid{0000-0002-8163-1191}

\author{Aditya Ukarande}
\affiliation{%
  \institution{University of Wisconsin - Madison}
  \city{Madison}
  \state{Wisconsin}
  \country{USA}
}
\email{ukarande@wisc.edu}
\orcid{0009-0009-2401-4902}

\author{Shilpa Mysore Srinivasa Murthy}
\affiliation{%
  \institution{University of Wisconsin - Madison}
  \city{Madison}
  \state{Wisconsin}
  \country{USA}
}
\email{ssrinivasamu@wisc.edu}

\author{Sahil Hassan}
\affiliation{%
  \institution{University of Arizona}
  \city{Tucson}
  \state{Arizona}
  \country{USA}
}
\email{sahilhassan@arizona.edu}
\orcid{0000-0002-4574-9555}

\author{Joshua Mack}
\affiliation{%
  \institution{University of Arizona}
  \city{Tucson}
  \state{Arizona}
  \country{USA}
}
\email{jmack2545@arizona.edu}
\orcid{0000-0003-1066-5578}

\author{Chaitali Chakrabarti}
\affiliation{%
  \institution{Arizona State University}
  \city{Phoenix}
  \state{Arizona}
  \country{USA}
}
\email{chaitali@asu.edu}
\orcid{0000-0002-9859-7778}

\author{Umit Ogras}
\affiliation{%
  \institution{University of Wisconsin - Madison}
  \city{Madison}
  \state{Wisconsin}
  \country{USA}
}
\email{uogras@wisc.edu}
\orcid{0000-0002-5045-5535}

\author{Ali Akoglu}
\affiliation{%
  \institution{University of Arizona}
  \city{Tucson}
  \state{Arizona}
  \country{USA}
}
\email{akoglu@arizona.edu}
\orcid{0000-0001-7982-8991}
\renewcommand{\shortauthors}{Gener et al.}

\begin{abstract} 
\new{Efficient memory management in heterogeneous systems is increasingly challenging due to diverse compute architectures (e.g., CPU, GPU, FPGA) and dynamic task mappings not known at compile time. Existing approaches often require programmers to manage data placement and transfers explicitly, or assume static mappings that limit portability and scalability. This paper introduces RIMMS (Runtime Integrated Memory Management System), a lightweight, runtime-managed, hardware-agnostic memory abstraction layer that decouples application development from low-level memory operations. RIMMS transparently tracks data locations, manages consistency, and supports efficient memory allocation across heterogeneous compute elements without requiring platform-specific tuning or code modifications. We integrate RIMMS into a baseline runtime and evaluate with complete radar signal processing applications across CPU+GPU and CPU+FPGA platforms. RIMMS delivers up to 2.43X speedup on GPU-based and 1.82X on FPGA-based systems over the baseline. 
Compared to IRIS, a recent heterogeneous runtime system, RIMMS achieves up to 3.08X speedup and matches the performance of native CUDA implementations while significantly reducing programming complexity. Despite operating at a higher abstraction level, RIMMS incurs only 1–2 cycles of overhead per memory management call, making it a low-cost solution. These results demonstrate RIMMS’s ability to deliver high performance and enhanced programmer productivity in dynamic, real-world heterogeneous environments.
}

\end{abstract}

\ccsdesc[500]{Software and its engineering~Runtime environments}
\ccsdesc[500]{Software and its engineering~Memory management}
\ccsdesc[500]{Computer systems organization~Heterogeneous (hybrid) systems}
\ccsdesc[400]{Computer systems organization~System on a chip}
\ccsdesc[100]{Computer systems organization~Real-time systems}

\keywords{Hardware-agnostic memory management, heterogeneous computing, runtime integration}


\maketitle

\section{Introduction}\label{sec:intro}
The limitations of traditional transistor scaling have sparked a renewed interest in computer architecture, leading to the emergence of domain-specific heterogeneous systems as a promising way forward. These systems integrate diverse processing elements (PEs), such as CPUs, GPUs, custom accelerators, and FPGAs, to overcome the shortcomings of general-purpose processors through specialized acceleration. While heterogeneous computing shows potential in enhancing energy efficiency and optimizing performance, it still cannot match the efficiency offered by Application-Specific Integrated Circuits (ASICs). Domain-Specific Architectures (DSAs) have emerged to address these limitations, narrowing the computational focus to specific domains to enable more efficient resource management and improve programmability~\cite{hennessy2019new,krishnakumar2023domain}. This approach has led to the development of Domain-Specific System-on-Chips (DSSoCs) with the aim of balancing flexibility and performance~\cite{TECS-Trireme_DSE-HW,DAC-LucaP15,NOCS-LucaP23,HCT-Survey-JeffV15}.

The increased heterogeneity envisioned in DSSoCs introduces significant challenges for application developers and system designers. For example, it is more difficult for developers to write applications that can effectively leverage all diverse resources available in such systems, as different PEs may require distinct programming models and optimization techniques. Furthermore, managing different compilation flows to deploy applications across heterogeneous components adds another layer of complexity for developers. Recent studies have addressed the aforementioned challenges in the form of ecosystems to enable productive and hardware-agnostic application development and deployment~\cite{TECSS-FARSI_DSE,TECSS-Stanford_CompDSE,CEDR2020HCW}, designing runtime systems~\cite{runtime-StarPU,runtime-TPDS-IRIS24,CEDR2023TECS,CEDR2024HCW}, intelligent scheduling frameworks~\cite{sched-HEFT_RT,sched-DAS,sched-HCW24-IRIS} capable of managing system resources while meeting quality of service requirements for each application sharing the system.

In heterogeneous systems, the variety and distribution of memory resources can be as diverse, if not more so, than the PEs themselves. To achieve optimal performance, programming interfaces must offer robust methods for representing and allocating memory across different device regions.  
The key challenge is creating memory allocation mechanisms that are aware of the memory hierarchy of the target compute platform, enabling data to be allocated directly to the most suitable memories.In a static deployment scenario, the programmer can hard code data flow into the application, knowing which PE executes each task in the application. In stark contrast, dynamic deployment scenarios face critical barriers. First, task-to-PE mappings are performed at runtime based on the dynamic system state. Hence, this knowledge cannot be hard coded at compile time. Designating a host CPU as the owner of the data simplifies data consistency and coherence among the PEs~\cite{IEEE_SPM_2009_MMovement1}. It enforces a flow where each PE receives its data for its assigned task from the host CPU, and the output is always sent back to the host CPU upon task completion~\cite{ICDE_2024_MMovement4}. While such a setup reduces the programmer's burden, it also requires data replication and redundant data transfers, particularly when successive tasks are executed on distinct PEs~\cite{PACT_2004_MMovement2}. The data flow overhead scales rapidly with the number of applications deployed on the system and becomes further complicated when applications exhibit a high degree of parallelism~\cite{Springer_SMA_2012_MMovement3,IEEE_TCS_2025_MMovement5}. 

This paper presents RIMMS, a novel Runtime Integrated Memory Management System, for heterogeneous and dynamic computing platforms where the task-to-PE mapping decisions are not fixed at compile time. We develop memory allocation techniques that are both \textit{platform-agnostic and adaptable to the specific resource management policies and hardware requirements of heterogeneous systems}. Particularly, we introduce hardware-agnostic memory allocation functions such that application developers do not have to be aware of the underlying hardware architecture. The compiler flow generates application binaries for the hardware-agnostic function prototypes since the inputs and outputs of these calls are known at compile time. These function definitions serve as hooks to enable the runtime system to track the location of the data and which PE updated it most recently. This information decreases the overall number of memory copies needed between PEs, resulting in less time spent on memory copies during the application lifetimes. RIMMS couples this approach with a lightweight bitset-based marking system that offers fast memory allocation and deallocation, and low memory usage due to its compact representation. To further improve allocation efficiency, we integrate an alternative memory marking approach that reduces computational overhead and accelerates allocation operations, when the marking system’s metadata is not constrained by limited memory space. Additionally, we introduce a mechanism for structured memory reuse, enabling efficient subdivision of allocated blocks without requiring additional allocations. This method maintains hardware-agnostic memory allocation and application development experience without incurring cycle overhead on the operations exclusive to the runtime system.

RIMMS is designed to operate on top of an existing runtime system and is not an extension of any particular one. In this work, we integrate RIMMS with the Compiler-integrated Extensible DSSoC Runtime (CEDR)~\cite{CEDR2020HCW,CEDR2023TECS,CEDR2023HCW}, which provides capabilities for dynamic task scheduling, application orchestration, and resource management in heterogeneous systems. While CEDR handles task execution, it does not manage data placement or movement across memory hierarchies. RIMMS addresses this gap by introducing a novel, runtime-integrated memory management and abstraction layer that tracks data locations, manages consistency, and enables hardware-agnostic memory allocation. As part of the RIMMS framework, we introduce a lightweight, platform-independent API that allows programmers to allocate and manage memory without needing to know where tasks will execute or how data will move between compute resources. This abstraction significantly reduces the complexity of programming for heterogeneous platforms. The runtime system makes dynamic task placement decisions based on system state, while RIMMS transparently ensures that data is allocated and transferred efficiently. Unlike CEDR, which focuses on task execution, RIMMS addresses the orthogonal and previously unhandled challenge of memory management in dynamic heterogeneous settings. By integrating compiler-inserted memory operations with runtime metadata tracking, RIMMS reduces redundant data transfers and simplifies application development. Although we demonstrate RIMMS using CEDR in this work, its design is portable and can be integrated with other runtime systems without requiring modifications to application code or platform-specific tuning.
To further validate its effectiveness, we use a reference application and compare the performance of RIMMS with that of IRIS~\cite{runtime-TPDS-IRIS24}, a modern open-source runtime, and native CUDA implementations. The evaluation shows that RIMMS consistently outperforms IRIS in total execution time by 1.35X to 3.08X, and tracks closely with low-level, hand-optimized CUDA code across a range of problem sizes, while incurring only 1-2 cycles per memory management call despite operating at a higher level of abstraction. These findings reinforce RIMMS's contribution as both a performance-efficient and programmer-friendly solution for portable memory management in heterogeneous systems.

We validate RIMMS and thoroughly evaluate its performance on both FPGA and GPU-based SoC platforms. These evaluations showcase its portability across different platforms and demonstrate the performance improvements achieved while running the target set of signal processing applications. In summary, the following are the main contributions of this work:

\begin{itemize}
    \item A runtime-integrated memory management system that enables memory allocation and data tracking in heterogeneous computing platforms.
    \item A hardware-agnostic memory allocation approach that abstracts the complexity of utilizing diverse memory architectures, allowing seamless application portability across different platforms.
    \item A compiler-assisted runtime mechanism that inserts function prototypes for memory operations, enabling the runtime to manage data locations and minimize redundant memory transfers without programmer involvement.
    \item A memory tracking and allocation strategy that explores the tradeoff between metadata overhead and efficiency: a lightweight bitset-based approach minimizes metadata for memory-limited systems, while a linked-list-based marking system reduces time spent on memory allocation by 2.55X for high-performance execution.
    \item Comprehensive validation and performance evaluation demonstrate that RIMMS reduces memory transfer overhead and improves execution efficiency on FPGA and GPU-based SoC platforms by up to 1.82X and 2.43X, respectively.
\end{itemize}

The rest of the paper is organized as follows. In Section~\ref{sec:background}, we review the related memory management systems and introduce the runtime framework. In Section~\ref{sec:mmu}, we introduce our proposed memory management approach and present its integration with the runtime system. We present heterogeneous SoC emulation and benchmark applications in Section~\ref{sec:expSetup}, followed by results in Section~\ref{sec:results}. Finally, in Section~\ref{sec:conclusion} we present our conclusions and future work.

\section{Related Work and Background}\label{sec:background}
Recent work has addressed memory management in heterogeneous systems, but often with limited heterogeneity~\cite{survey-Hetero-MMU}. For instance, the approach in~\cite{MM-Multi-Core-MICRO23} focuses solely on heterogeneous CPU clusters and lacks the ability to manage memory across more diverse PEs. Similarly, CPU-FPGA systems~\cite{MM-CPU-FPGA-ICCD13,MM-CPU-FPGA-JETCS17} leverage FPGA-specific logic, such as data reuse buffers within the programmable fabric. However, they are highly specialized for FPGA-centric architectures, making them difficult to generalize to other forms of heterogeneity. Various memory optimization techniques have been proposed in the context of CPU-GPU heterogeneity. Studies in~\cite{MM-CPU-GPU-SIGARCH14},~\cite{MM-CPU-GPU-HPCA14}, and~\cite{MM-CPU-GPU-MICRO17} modify TLBs and page tables within GPUs to enhance performance. Meanwhile, in~\cite{MM-CPU-GPU-ASPLOS20}, authors introduce a batch-aware GPU runtime solution, incorporating both hardware and software changes to reduce the frequency of page faults in GPU memory. In~\cite{MM-CPU-GPU-IEEEAccess24}, another CPU-GPU-based system minimizes memory consumption and management overhead in legacy GPU applications by optimizing memory usage between host and device allocations. However, the scope of these approaches remains confined to GPU architectures.

In summary, prior approaches remain constrained by their focus on specific subsystems, such as CPUs, FPGAs, or GPUs, and do not address the broader challenges of managing memory across fully heterogeneous systems with diverse PEs. Consequently, none of these solutions can be directly applied to environments with more complex heterogeneity, where a unified memory management strategy is required. Hence, in this work, we adopt the default memory management scheme of the used runtime system and compare our results against that baseline.


We choose the open-source Compiler-integrated Extensible DSSoC Runtime (CEDR)~\cite{CEDR2020HCW,CEDR2023TECS,CEDR2023HCW} framework as our runtime system since it
offers three advantages:
(1) a unified 
framework capable of handling simultaneous application executions, (2) flexibility to accommodate varying heterogeneity and workload compositions, and (3) portability across off-the-shelf heterogeneous platforms. CEDR's application programming interface (API) model also enables application deployment across different systems without requiring knowledge of the underlying hardware. Additionally, its integrated scheduler enables the execution of multiple applications while optimizing resource sharing within the system. This environment allows us to integrate our memory management approach and extend its functionalities. \textit{While we implement our approach within CEDR, the underlying techniques are general and could be adapted to other runtime frameworks.}

\begin{lstlisting}[
            language=cpp,
            firstnumber=1,
            basicstyle=\scriptsize\ttfamily,
            label={lst:cuda_copy},
            caption={Application implementation example using CUDA},
            float=tp,
            escapechar=\%
        ]
#include <cuda_runtime.h>
...
// Instantializations
cufftComplex *host_input, *device_input, *host_output, *device_output;
bool FORWARD = true;
// Allocations
host_input = (cufftComplex*) malloc(M * N * sizeof(cufftComplex));
host_output = (cufftComplex*) malloc(M * N * sizeof(cufftComplex));
cudaMalloc((void**)&device_input, M * N * sizeof(cufftComplex));
cudaMalloc((void**)&device_output, M * N * sizeof(cufftComplex));
...
// Explicit memory copy
cudaMemcpy(device_input, host_input, M * N * sizeof(cufftComplex), cudaMemcpyHostToDevice);
// Execution of M FFTs of size N
for (int i = 0; i < M; i++){ // FFT function abstraction
    cuda_fft_wrapper(&device_input[i * N], &device_output[i * N], N, FORWARD);
}
cudaDeviceSynchronize();
// Explicit memory copy
cudaMemcpy(host_output, device_output, M * N * sizeof(cufftComplex), cudaMemcpyDeviceToHost);
...
// Deallocations
cudaFree(device_input); cudaFree(device_output);
free(host_input); free(host_output);
\end{lstlisting}

CUDA~\cite{CUDA_2008} and OpenMP~\cite{OpenMP_1998} provide mechanisms for managing memory transfers and executing tasks on accelerators. However, both rely on static resource allocations, making them less adaptable to dynamic heterogeneous systems. As shown in Listing~\ref{lst:cuda_copy}, CUDA requires explicit memory allocation using \emph{cudaMalloc} (lines 9 and 10) and data movement via \emph{cudaMemcpy} (lines 13 and 20) to transfer data between host and device. While zero-copy memory and unified memory (\emph{cudaMallocManaged}) are available, explicit memory copies remain the most efficient option. Similarly, Listing~\ref{lst:openmp_copy} illustrates how OpenMP offloading uses \emph{\#pragma omp target} directives with explicit \emph{map(to/from)} clauses (lines 11 and 19), to manage data transfers between the host and the PE -- typically a GPU but potentially any supported device. In both CUDA and OpenMP, execution is statically mapped to a single device, requiring explicit data movement. 
In contrast, the proposed system tracks data ownership, allowing it to dynamically determine which resource last modified the data. Task execution is managed through CEDR APIs, which handle dynamic task dispatching to available resources and abstract memory management. By eliminating explicit memory copies and supporting dynamic scheduling, the proposed approach is better suited for heterogeneous environments with dynamic resource allocations.

\begin{lstlisting}[
            language=cpp,
            firstnumber=1,
            basicstyle=\scriptsize\ttfamily,
            label={lst:openmp_copy},
            caption={Application implementation example using OpenMP},
            float=tp,
            escapechar=\%
        ]
#include <omp.h>
...
// Instantializations
complex<float> *input, *output;
bool FORWARD = true;
// Allocations
input = (complex<float>*) malloc(M * N * sizeof(complex<float>));
output = (complex<float>*) malloc(M * N * sizeof(complex<float>));
...
// Explicit memory copy
#pragma omp target enter data map(to: input[:M*N])

// Execution of M FFTs of size N
#pragma omp target teams distribute parallel for
for (int i = 0; i < M; i++){ // FFT function abstraction
    fft_wrapper(&input[i * N], &output[i * N], N, FORWARD);
}
// Explicit memory copy
#pragma omp target exit data map(from: output[:M*N])
...
// Deallocations
free(input); free(output);
\end{lstlisting}

Runtimes like IRIS~\cite{runtime-TPDS-IRIS24} and StarPU~\cite{runtime-StarPU} provide mechanisms for managing resource allocation in heterogeneous environments. 
IRIS, in particular, offers well-designed APIs that abstract low-level hardware details and support diverse backends for task execution--including CUDA, OpenCL, and HIP--making it a practical choice for targeting multiple accelerators within a unified runtime system, which is primarily developed for HPC-like systems. Its task-based programming model promotes portability and allows developers to express parallelism at a higher level. Listing~\ref{lst:iris_copy} illustrates a brief snippet of host C code for the same FFT implementation using IRIS, which manages memory through its own data structures and functions. However, like CUDA and OpenMP, IRIS requires developers to manually invoke functions such as \emph{iris\_task\_d2h} and \emph{iris\_task\_h2d} to move data before and after task execution (lines 15 and 19). \textit{In contrast, this work eliminates the need for these explicit calls, enabling seamless memory management and simplifying application development.}
\begin{lstlisting}[
            language=cpp,
            firstnumber=1,
            basicstyle=\scriptsize\ttfamily,
            label={lst:iris_copy},
            caption={Application implementation example for IRIS~\cite{runtime-TPDS-IRIS24}},
            float=tp,
            escapechar=\%
        ]
#include <iris/iris.h>
...
// Instantializations
float *input, *output;
iris_mem mem_input, mem_output;
iris_task task;
bool FORWARD = true;
// Allocations
input  = (float*) malloc(N * sizeof(float));
output = (float*) malloc(N * sizeof(float));
iris_mem_create(N * sizeof(float), &mem_input);
iris_mem_create(N * sizeof(float), &mem_output);
...
// Explicit memory copy from Host to Device
iris_task_h2d(task, mem_intput, 0, N * sizeof(float), input);
// Execution of M FFTs of size N
/* IRIS setup for setting up M FFTs of size N -- omitted */
// Explicit memory copy from Device to Host
iris_task_d2h(task, mem_output, 0, N * sizeof(float), output);
iris_task_submit(task, iris_greedy, NULL, 1);
iris_synchronize();
...
// Deallocations
free(input); free(output);
iris_mem_release(mem_input); iris_mem_release(mem_output);
\end{lstlisting}

\section{Memory Management Unit}\label{sec:mmu}
In this section, we introduce the proposed memory management-related API calls and apply compile-time modifications to the application, ensuring seamless integration of these changes. To quantify the benefits of the proposed widely applicable memory management approach, we use the data flow management approach supported by CEDR as a reference. 

\subsection{Reference Implementation}
\begin{figure}[t]
    \centering
    \begin{subfigure}{0.45\textwidth}
        \centering
        \includegraphics[width=0.9\textwidth]{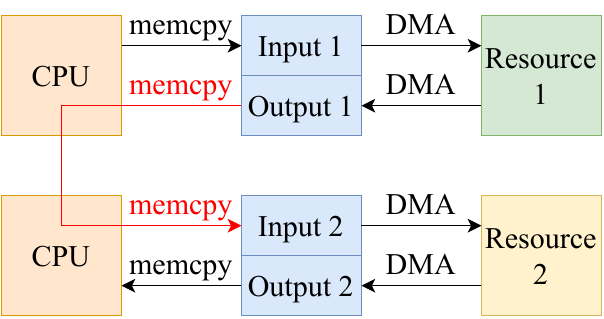}
        \caption{With redundant memcpy}
    \end{subfigure}
    \begin{subfigure}{0.45\textwidth}
        \centering
        \includegraphics[width=0.9\textwidth]{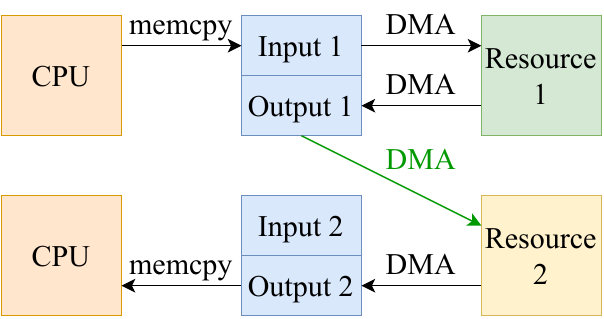}
        \caption{Without redundant memcpy}
    \end{subfigure}
    \caption{Scenario with (a) redundant memory copies on data operated by two PE types and (b) elimination of redundancy with direct data flow between PEs.}
    \Description{To be updated after acceptance.}
    \label{fig:redundant-memcpy}
\end{figure}

In the current mainstream approaches as in CPU-GPU coupled systems, data is typically owned by the host CPU. This means that whenever data is processed by a resource, it must first be copied from the host CPU's memory space to the resource's associated memory space. After the resource completes execution, the data is copied back to the host CPU's memory space to ensure the host always has an up-to-date copy. While this flow guarantees data consistency, it also introduces redundant data transfers when the host CPU does not need to access or use the data between consecutive executions on the same resource. We refer to this type of data management as the reference implementation during our analysis. Figure~\ref{fig:redundant-memcpy}(a) illustrates a scenario where \emph{Resource 1} sends data to the \emph{Resource 2} via CPU. It involves redundant data copy operations that can be avoided via direct resource-to-resource data flow, as illustrated with Figure~\ref{fig:redundant-memcpy}(b). At the same time, the Direct Memory Access (DMA) engines can be configured so that, instead of \emph{Resource 2} reading from its separate \emph{Input 2} buffer, it can directly read from \emph{Resource 1}'s \emph{Output 1} buffer.

\subsection{The Proposed RIMMS Approach}
If the producer and consumer relationships are known at design time, data flow can be managed by the tasks themselves. Although guaranteed to realize an optimal data flow, this setup has a key drawback. In order to make all successor and predecessor task information available, the application developer needs to provide a directed acyclic graph (DAG) with the application. This, in turn, increases the complexity of application development. Furthermore, when the task-to-PE mapping decisions change dynamically, the assumption of a certain task being executed on a single type of resource is no longer valid. Therefore, static decision-based memory management will not result in optimal performance. Instead of having tasks track their execution resources, our approach is to shift this responsibility to the data itself, allowing the data to retain information about the last resource that modified it. 

We introduce hardware-agnostic memory management functions and protocols, creating a generalizable and more user-friendly memory management system. This combination results in an integrated system where compile-time hardware-agnostic memory management calls become accessible to the runtime system. This system handles data allocation at compile-time and automatically manages data flow at runtime without requiring users to embed data flow management directives specific to the target hardware architecture and memory hierarchy into the application. Handling data flow management under the hood is beneficial in two key ways. First, the proposed approach hides the complexity of data flow management for systems with a high degree of heterogeneity. 
Secondly, and equally important, it enables portability across heterogeneous platforms. 

\begin{figure}[t]
    \centering
    \includegraphics[width=\textwidth]{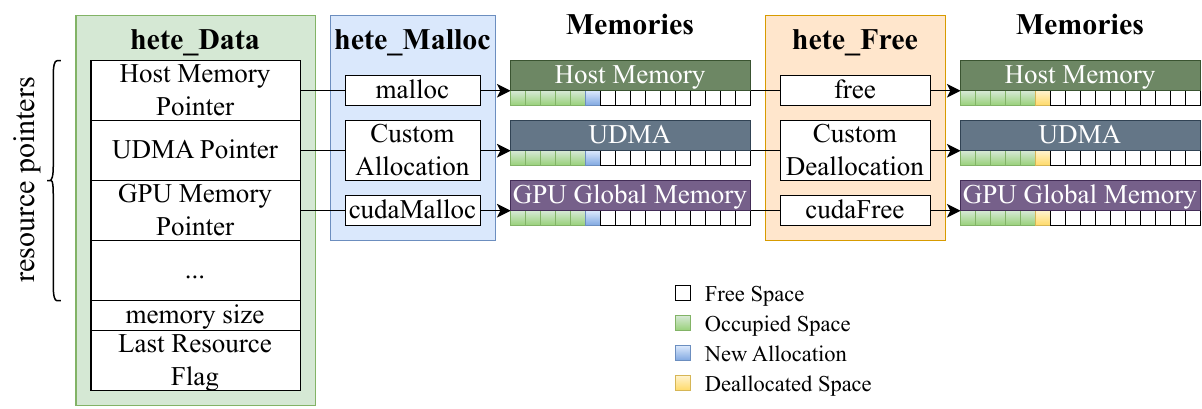}
    \caption{Contents of \cedrdata~data structure and underlying flow of new \cedrmalloc~and \cedrfree~APIs.}
    \Description{To be updated after acceptance.}
    \label{fig:cedr_data}
\end{figure}

We introduce a new data structure, \cedrdata, which maintains pointers to different memory regions (\emph{resource pointers}) and tracks the last location where the data is updated (\emph{last resource flag}). Additionally, we develop three new API calls, including \cedrmalloc~for memory allocation, \cedrfree~for memory deallocation, and \cedrsync~for synchronizing memory between the host CPU and other resources. The following subsections 
describe our design approach for each. The new data structure and the APIs are illustrated in Figure~\ref{fig:cedr_data}.

\subsubsection{Hardware-Agnostic Memory Management Functions}\label{subsec:mmfunctions}

The \cedrmalloc~function provides users with a pointer to the \cedrdata~structure, offering transparent access to the \emph{data} field, which resides in the host CPU’s memory. This design allows developers to interact with the data without worrying about underlying hardware-specific details. The API calls internally manage fields such as \emph{resource pointers}, which refer to memory locations on specific resources, and the \emph{last resource flag}, which keeps track of which pointer holds the valid data. Although these fields are visible to the application developer, they are explicitly managed by the API calls and do not require user involvement. 
To maintain hardware-agnostic design, we have modified the inputs and outputs of existing API calls in CEDR to use the \cedrdata~type. This change allows the compiler to set up platform-specific memory allocation at compile time, ensuring that runtime execution occurs seamlessly on any target hardware. Users only need to specify the \emph{memory size} in bytes, similar to a standard C/C++ \emph{malloc} call, while the appropriate memory for the resource is allocated automatically at runtime under the hood, as shown in Figure~\ref{fig:cedr_data}.

We also introduce a new API, \cedrfree, to complement \cedrmalloc. Similar to the relationship between \emph{malloc} and \emph{free}, \cedrfree~deallocates memory by freeing all resource pointers associated with a given \cedrdata~structure. This deallocation fully releases all the memory previously allocated by \cedrmalloc, making memory space available for future allocations.
By handling both CPU and resource-specific memory regions, the hardware-agnostic \cedrfree~functional call seamlessly handles memory deallocation on the specific memory of the target hardware architecture (shown in Figure~\ref{fig:cedr_data}) where the data resides at runtime. This approach also simplifies the deallocation process across different hardware platforms while maintaining a consistent and user-friendly interface.

The third API introduced is \cedrsync. This function is required when the user needs to read from or write into the specific data allocated through \cedrmalloc~within the application without using another API call. If an API call uses a \cedrdata~and it requires modification on the host CPU on the application side, \cedrsync~ensures that the data in the host CPU's memory is synchronized with its most up-to-date version, which may reside in resource-specific memory. 

\subsubsection{Memory Management Protocols}\label{subsec:mmprotocols}

We utilize a lightweight bitset-based marking system for memory allocation and deallocation as a heap management scheme. This method's compact representation minimizes the memory overhead of maintaining heap metadata. We divide the resource memory regions into blocks, using bitsets to mark blocks as \textit{used} during allocation and clearing them during deallocation. When allocating memory, the runtime system searches for consecutive blocks whose total size, in terms of bytes, is at least equal to the requested number of bytes. If there is not enough space for allocation, the runtime system is terminated. While block sizes can be adjusted as needed, they remain fixed during CEDR's runtime. The memory footprint of this approach is only 1 bit per block for tracking its usage.

While the bitset-based marking system minimizes metadata memory overhead, it is computationally expensive due to its exhaustive search for contiguous free blocks. To mitigate this overhead, if the metadata is not stored in the limited-resource memory, we introduce a next-fit-based (NF-based) marking system using a linked-list metadata structure, which optimizes allocation by maintaining a rolling search position, while increasing the memory footprint to approximately 17 bytes per metadata entry. Initially, the entire resource memory is marked as unused. During allocation, the search begins from the last allocated position and selects the next available space that meets or exceeds the requested size. Once a suitable block is found, it is split into two: the first segment, sized precisely to the request, is marked as used, while the remaining portion remains unused as a separate block. The search position is then updated to this unused block for future allocations. During deallocation, a block is marked as unused and merged with adjacent free blocks, if any, to reduce fragmentation. While this approach may increase fragmentation compared to the bitset-based method, it significantly improves computational efficiency. Additionally, it imposes no fixed block size constraints, allowing flexible allocations of arbitrary sizes.

We update the resource-specific function calls, functions that APIs fall back to once assigned to a resource, within CEDR in two ways. First, each function verifies the last modification location of the data it receives by checking the \emph{last resource flag}. If the data is in the host CPU memory, the function copies it to the resource-specific memory; if it is in resource-specific memory, the function takes no action regarding data movement. Functions perform this check for all inputs. Second, as a final step, the resource-specific function updates the \emph{last resource flag} to indicate that the data was last modified during execution on the associated resource. If a function has more than one output, it repeats this process for each output. Since CEDR, by design, forces parallelism at the API level rather than allowing multiple resources to process the same data simultaneously, each API call is strictly assigned to a single resource for execution. RIMMS inherits this and guarantees clear ownership of input and output data per API. The \emph{last resource flag} corresponds unambiguously to the single resource responsible for processing a given data, eliminating race conditions or conflicts over this flag by design.

The protocols described in this section have been adapted for platforms such as SoC-based FPGAs, which require data to be placed in physically contiguous memory before being transferred to accelerators. To facilitate this, Unified DMAs (UDMAs) are used in this work, necessitating custom allocation and deallocation schemes. Unlike GPUs, which offer built-in memory management mechanisms such as \emph{cudaMalloc} and \emph{cudaFree}, FPGAs lack native support for memory allocation and deallocation, particularly for UDMA usage. While it is not possible for UDMAs on FPGAs, whenever possible, we leverage existing memory management mechanisms when users invoke the hardware-agnostic APIs proposed in Section~\ref{subsec:mmfunctions}. For example, when an application calls \cedrmalloc, it internally utilizes \emph{cudaMalloc} on NVIDIA-based GPUs.

\subsubsection{Data Fragmentation}\label{subsec:mmfragment}
Developers often allocate a large one-dimensional array and reference portions of it as if it were a two-dimensional array through indexing. This approach allows memory to be allocated only once while still accessing structured data efficiently. For example, when executing $M$ instances of an $N$-point FFT, a developer can allocate a single $M \times N$ array and index every $N$ element separately for each FFT. However, in the \cedrdata~structure, multiple types of data pointers must be managed, making direct indexing infeasible while still tracking the \emph{last resource flag} for each data section. As a result, using \cedrdata~for the same example requires calling \cedrmalloc~$M$ times, once per FFT. Including both input and output allocations, this results in $2 \times M$ allocations, whereas traditional allocation without \cedrdata~would only require $2$. This introduces a significant overhead in the allocation (results will be shown in Section~\ref{subsec:allocation_overhead_PD}) for real applications when using RIMMS and diminishes the improvement gained from eliminating redundant memory transfers. To address this issue, we introduced a \emph{fragment} function for \cedrdata, allowing an already allocated memory block to be subdivided into multiple regions, each maintaining its own data pointers and \emph{last resource flag}. With this approach, developers can allocate a single $M \times N$ block using \cedrmalloc~and invoke the \emph{fragment} function to create internal fragmented \cedrdata~pointers within the allocated \cedrdata. Since this operation does not modify resource memory mapping, it eliminates the overhead of searching for suitable memory locations during allocation, significantly accelerating the creation of $M$ data regions of size $N$. To further improve usability, we overloaded the indexing operation for \cedrdata. If indexing occurs after a \emph{fragment} operation, the index $i$ directly references the $i^{th}$ fragment in the data structure, simplifying accessing the data and application development process.
This fragmentation operation has linear execution complexity, that is, $O(n)$, where $n$ is the number of fragments requested by the user.

\subsubsection{Implementation Example}

\begin{lstlisting}[
            language=cpp,
            firstnumber=1,
            basicstyle=\scriptsize\ttfamily,
            label={lst:cedr_copy},
            caption={Application implementation example using RIMMS},
            float=tp,
            escapechar=\%
        ]
#include <cedr.h>
...
// Instantializations
$hete_Data$ *input, *output;
bool FORWARD = true;
// Allocations
input = $hete_Malloc$(M * N * sizeof(complex<float>));
output = $hete_Malloc$(M * N * sizeof(complex<float>));
input->$fragment$(N * sizeof(complex<float>));
output->$fragment$(N * sizeof(complex<float>));
...
// Execution of M FFTs of size N
for (int i = 0; i < M; i++){
    CEDR_FFT(&input[i], &output[i], N, FORWARD);
}
$hete_Sync$(output); // Only needed if output will be used later in a CPU-only operation
printResults(output);
...
// Deallocations
$hete_Free$(input); $hete_Free$(output);
\end{lstlisting}

\begin{figure}[t]
    \centering
    \includegraphics[width=\textwidth]{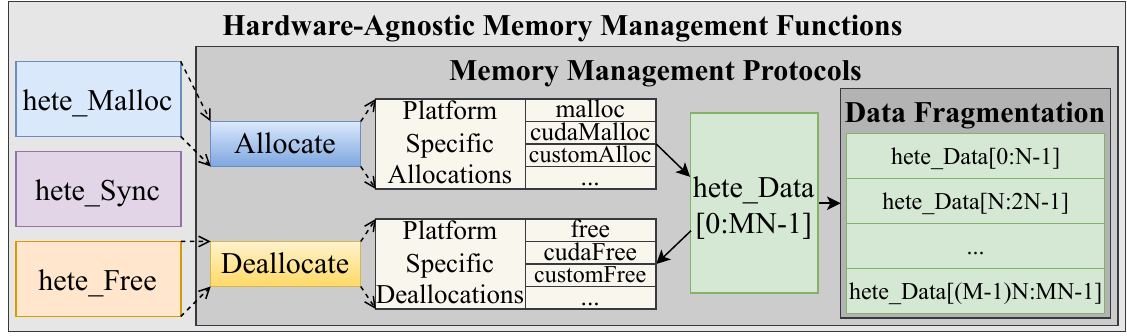}
    \caption{Overview of the hardware-agnostic memory management functions, memory management protocols, and data fragmentation.}
    \Description{To be updated after acceptance.}
    \label{fig:RIMMS_flow}
\end{figure}

Figure~\ref{fig:RIMMS_flow} illustrates the interactions between hardware-agnostic memory management functions (Section~\ref{subsec:mmfunctions}), memory management protocols (Section~\ref{subsec:mmprotocols}), and data fragmentation approach (Section~\ref{subsec:mmfragment}). Compared to Listing~\ref{lst:cuda_copy} and Listing~\ref{lst:openmp_copy}, the proposed implementation, shown in Listing~\ref{lst:cedr_copy}, introduces a simpler approach to memory management. It utilizes \cedrmalloc~(lines 7 and 8) for data allocation, while the \cedrdata~structure (line 4) maintains the \emph{last resource flag} to track data locality across heterogeneous resources. To enable indexed access (line 14) during application development, the data structure is fragmented using the \emph{fragment} function (lines 9 and 10). The \cedrsync~function can be used by the programmer to maintain consistency when application code accesses data outside the defined API boundaries, especially from the CPU. For example, if a compute resource modifies \emph{output} (line 14) and the CPU later reads that data directly (line 17)--i.e., not via another API call--\cedrsync~(line 16) ensures coherence by updating the memory state. Task execution is managed through \emph{CEDR\_FFT}, which dynamically assigns tasks to available resources. With the updated API, memory management for inputs and outputs is fully abstracted, further reducing developer overhead. Designed for environments with dynamic resource allocation, RIMMS eliminates the need for explicit memory transfers while supporting efficient task scheduling. 

\subsubsection{\new{System Compatibility}}
RIMMS is designed to operate transparently within the coherence and memory management policies of the host system. For standard memory allocations (e.g., those performed via malloc), RIMMS inherits the system's default behavior for memory consistency and NUMA awareness. Since it does not override or alter allocation mechanisms, RIMMS remains completely compatible with diverse NUMA topologies and memory coherence protocols provided by the operating system and hardware. For FPGA-based memory regions (e.g., UDMA buffers), RIMMS uses a kernel-level allocator that reserves physically contiguous memory pages and maps them into user space. These buffers are allocated from DDR memory and maintain coherence with other memory regions according to the platform's hardware protocols. On the GPU side, memory is explicitly allocated in global device memory through vendor APIs (e.g., CUDA). Host-device coherence is maintained using platform-specific synchronization mechanisms (e.g., cudaMemcpy). RIMMS respects these protocols and does not introduce additional coherence assumptions or constraints. In short, while RIMMS does not directly manage or enforce coherence policies, it remains agnostic to the specific protocol. It is fully compatible with systems featuring various coherence models and NUMA configurations,  provided the hardware and operating system support them. RIMMS also relies on the security mechanisms provided by the underlying system, including operating system (OS) protections and hardware-enforced memory isolation. Access controls, secure memory regions, and isolation policies enforced by the OS or runtime environment remain fully effective when RIMMS is in use. Since RIMMS operates entirely at the user-level runtime without bypassing these protections, it does not introduce additional vulnerabilities or compromise secure data handling other than those introduced by the OS or the runtime.

\section{Heterogeneous SoC Emulation and Setup}\label{sec:expSetup}
\subsection{Emulation and SoC Platforms}
We emulate a heterogeneous SoC using the Xilinx Zynq Ultrascale+ ZCU102 development board~\cite{ZCU102}. The system consists of four ARM CPU cores, each running at 1.2 GHz, two Fast Fourier Transform (FFT) accelerators implemented with the Xilinx FFT IP, and a pointwise vector operation (ZIP) accelerator implemented using HLS. Both FFT and ZIP work with complex float numbers. The FFT and ZIP accelerators operate at 300 MHz and utilize AXI4-Stream~\cite{AXI4} with DMA to manage data transfers. A 64 MiB UDMA buffer is used as the resource memory described in Section~\ref{subsec:mmprotocols}. To demonstrate the versatility of the proposed memory management, we leverage a GPU-based SoC, Jetson Xavier AGX~\cite{Jetson} platform with a 512-core Volta GPU running at 1.3GHz, which supports running FFT and ZIP API calls and eight ARM CPU cores, each running at 2.3GHz. While an ideal evaluation would include concurrent use of CPU, GPU, and FPGA on a unified platform, such a configuration is not currently available to us. Instead, we evaluate RIMMS on two representative heterogeneous setups (CPU+GPU and CPU+FPGA), capturing diverse memory and execution behaviors. These configurations, combined with complete radar signal processing workloads (detailed in Section 4.3) from the CEDR framework, stress RIMMS’s ability to manage memory allocation, consistency, and data movement across dynamic execution paths. Although prior work has explored full-system, multi-application execution involving all compute elements~\cite{CEDR2023TECS}, our focus here is to isolate and evaluate improvements specifically related to memory abstraction, tracking, and allocation under dynamic runtime task mapping. We also extend our evaluation to the IRIS~\cite{ runtime-TPDS-IRIS24} runtime by adapting one of its workloads to match the complexity of our use cases, providing a broader and more meaningful baseline for comparison.

\begin{figure}[t]
    \centering
    \begin{subfigure}{0.33\textwidth}
        \centering
        \includegraphics[width=0.9\textwidth]{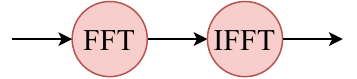}
        \vspace{15pt} 
        \caption{2FFT}
    \end{subfigure}
    \begin{subfigure}{0.33\textwidth}
        \centering
        \includegraphics[width=\textwidth]{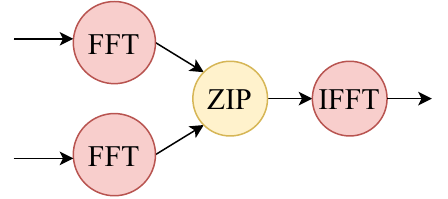}
        \caption{2FZF}
    \end{subfigure}
    \begin{subfigure}{0.32\textwidth}
        \centering
        \includegraphics[width=0.9\textwidth]{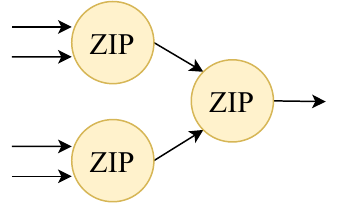}
        \caption{\new{3ZIP}}
    \end{subfigure}
    \caption{Representative signal processing chains for validation: (a) FFT to IFFT flow, (b) Two FFTs to ZIP to IFFT flow, and (c) Two ZIP to another ZIP flow}.
    \Description{To be updated after acceptance.}
    \label{fig:synthetic-application}
\end{figure}

\subsection{Test Applications and Experiments}
We use three distinct signal processing chains, illustrated in Figure~\ref{fig:synthetic-application}, to validate our approach and study the performance of RIMMS as a function of data size. The first reference application (2FFT) involves FFT followed by IFFT. It conceptually represents a common data flow where two accelerated functions run back-to-back. The second application (2FZF) involves two concurrent FFTs feeding their results to the ZIP multiplication, whose output is then fed into the IFFT. The third application (3ZIP) consists of a chain of three ZIP operations, where the final ZIP depends on the outputs of the first two. These three data flow structures with FFT and ZIP operations are selected since they are commonly observed in a wide range of radar and signal processing applications. We scale the number of samples for the FFT and ZIP kernels from 64 to 2,048 to represent workloads seen in modulation classification, energy detection, channel estimation, and beamforming classes of key kernels from the millimeter wave, 5G, and spectrum sensing applications~\cite{app-2023-AS,app-2023-IEEEAccess,app-nasser2021spectrum,app-van2021millimeter,app-vannithamby2017towards5G}.

\subsection{Experimental Setup}
To fairly evaluate memory management in heterogeneous systems, benchmarks must exhibit behaviors that stress memory allocation, inter-kernel data movement, synchronization, and dynamic reuse. These include frequent and dynamic memory allocations, tightly coupled task dependencies, and staged execution patterns across heterogeneous compute elements. Such characteristics are essential to meaningfully exercise and evaluate the benefits of a memory management framework like RIMMS. Commonly used heterogeneous benchmark suites, such as MiBench~\cite{MiBench}, Rodinia~\cite{Rodinia}, and HeteroBench~\cite{HeteroBench}, fall short in this regard due to limited parallelism, minimal memory reuse, and lack of complex inter-kernel data dependencies. Instead, to demonstrate the robustness of RIMMS, we present end-to-end compilation and deployment of real-world applications provided by the CEDR framework~\cite{CEDR2023TECS}, including Radar Correlator (RC), Pulse Doppler (PD), and Synthetic Aperture Radar (SAR). We repeat each reference application 10,000 times and find the average execution time. RC simulates radar pulse detection by measuring the time delay between transmitted and received pulses using three 256-point FFTs at a rate of 1,000 samples per second. PD emits short radar pulses and uses frequency shifts to determine object distance and velocity. It consists of four phases: the first with 256 parallel 128-point FFTs, the second with 128 parallel ZIPs, the third with 128 parallel 128-point FFTs, and finally the fourth with data rearrangement operations followed by 128 parallel 128-point FFTs. SAR is a data acquisition method for 3D surface reconstruction, utilizing 1,537 FFTs and 768 ZIPs. Its workflow includes phases of 512-way and 256-way parallel flow of FFT feeding into ZIP and ZIP feeding into FFT (FZF) for sample sizes of 256 and 512, respectively. In summary, these workloads feature rich inter-kernel dependencies, frequent memory reuse, and phased execution flows (e.g., FFT $\rightarrow$ ZIP $\rightarrow$ FFT). These characteristics create diverse, highly dynamic execution patterns, representative of workloads targeting DSSoC environments, and thus offer a more rigorous and realistic evaluation for RIMMS. We integrate RIMMS with CEDR and use CEDR as the reference in Section~\ref{sec:results} to evaluate the effectiveness of our approach based on end-to-end application execution times.

\section{Experimental Results}\label{sec:results}
\subsection{Experiments with the 2FFT Data Flow}

\begin{figure}[t]
    \centering
    \begin{subfigure}{0.48\textwidth}
        \centering
        \includegraphics[width=\textwidth]{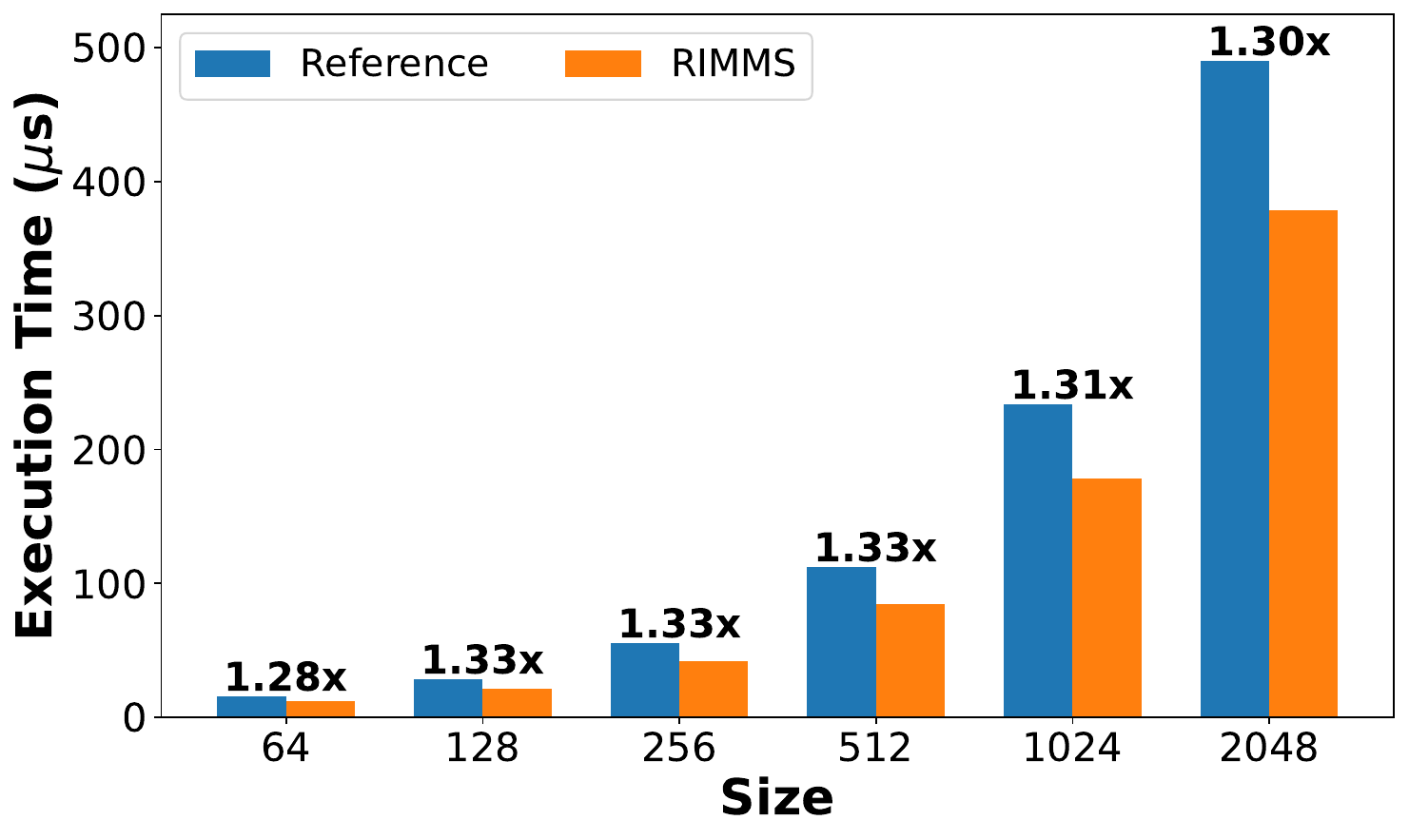}
        \caption{First FFT runs on CPU, second runs on accelerator}
    \end{subfigure}
    \begin{subfigure}{0.48\textwidth}
        \centering
        \includegraphics[width=\textwidth]{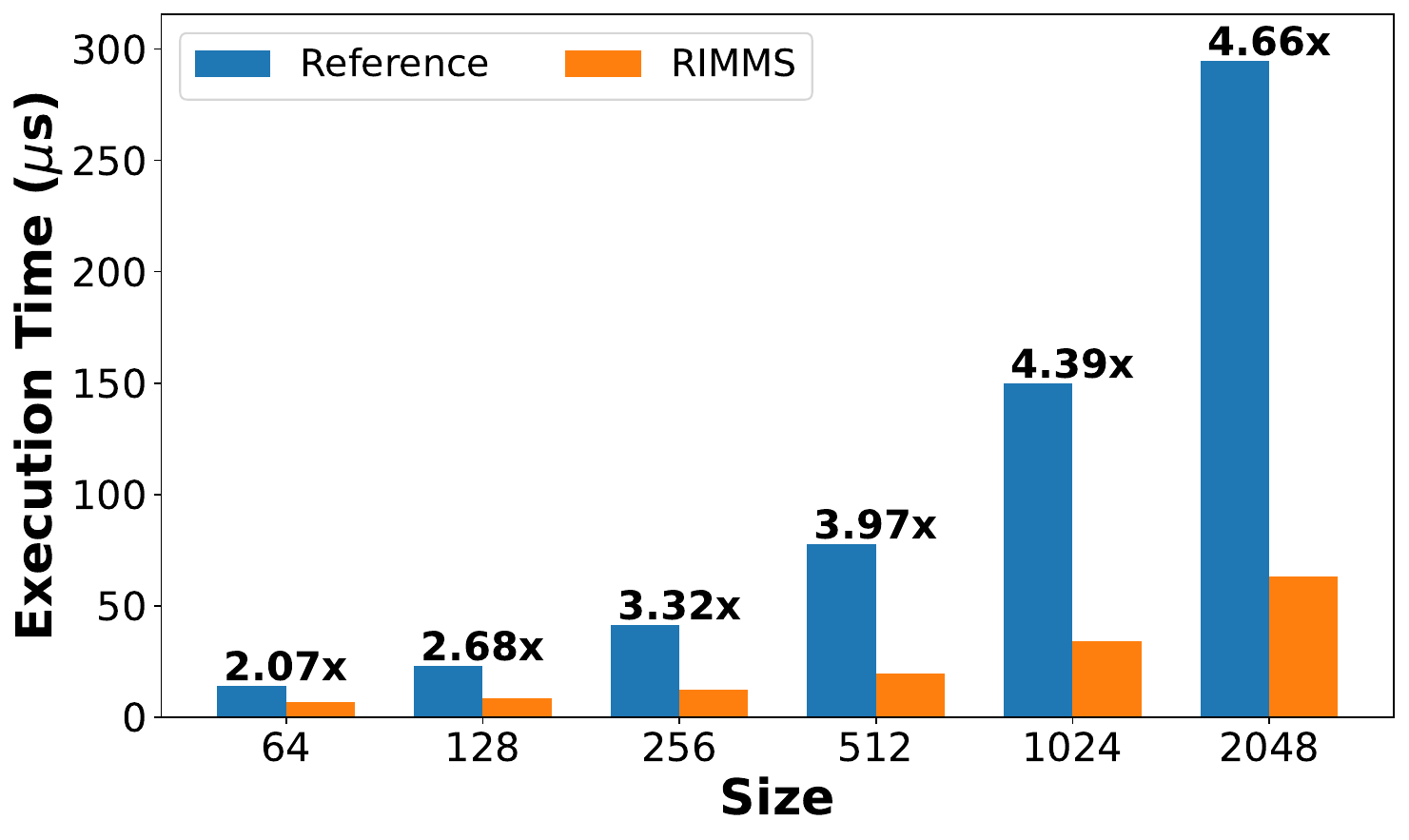}
        \caption{Both FFTs run on FFT accelerators}
    \end{subfigure}
    \caption{Execution time of 2FFT as a function of FFT size on ZCU102 using the reference system and RIMMS.}
    \Description{To be updated after acceptance.}
    \label{fig:zcu102-synthetic-2fft}
\end{figure}

\begin{figure}[t]
    \centering
    \begin{subfigure}{0.48\textwidth}
        \centering
        \includegraphics[width=\textwidth]{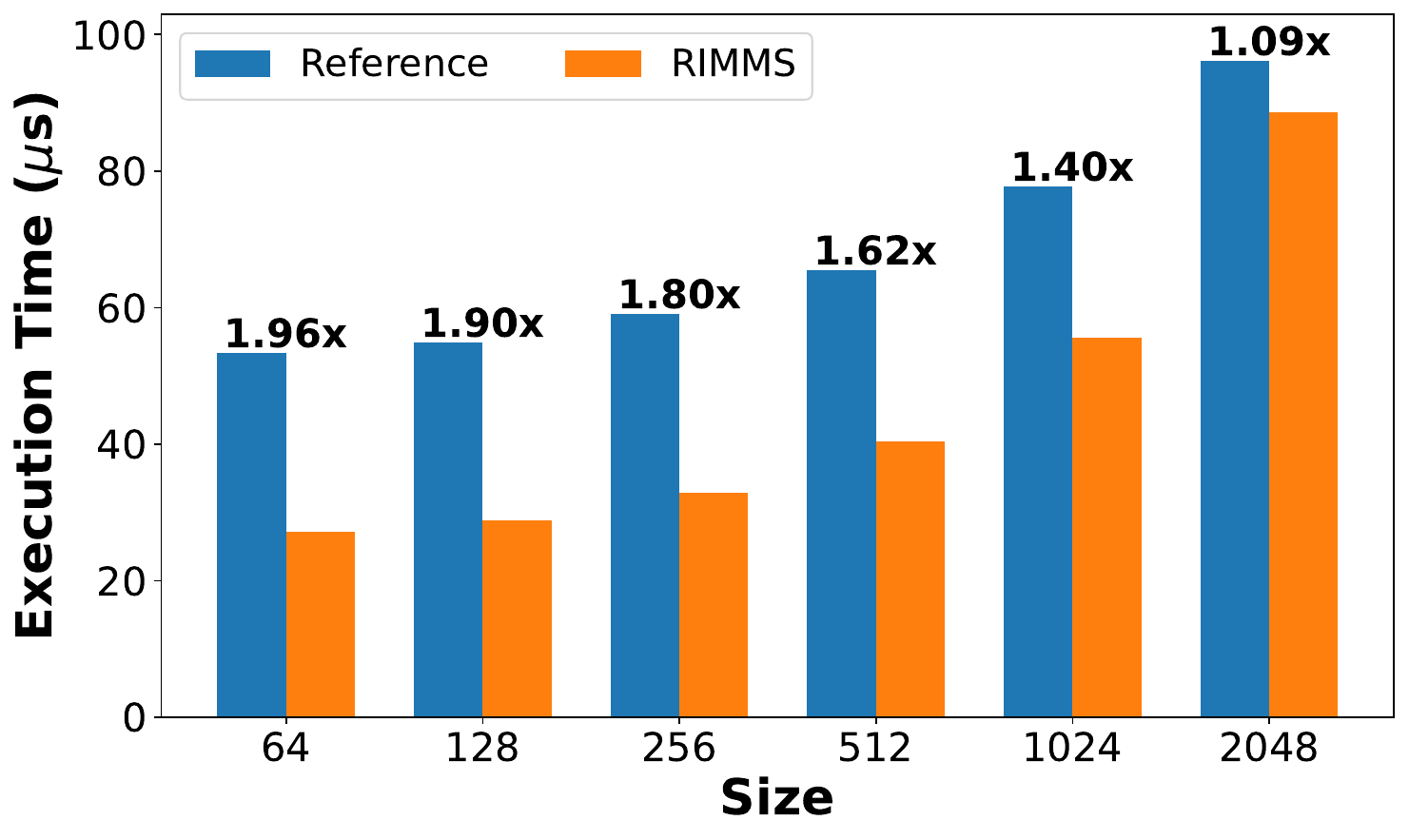}
        \caption{First FFT runs on CPU, second runs on GPU}
    \end{subfigure}
    \begin{subfigure}{0.48\textwidth}
        \centering
        \includegraphics[width=\textwidth]{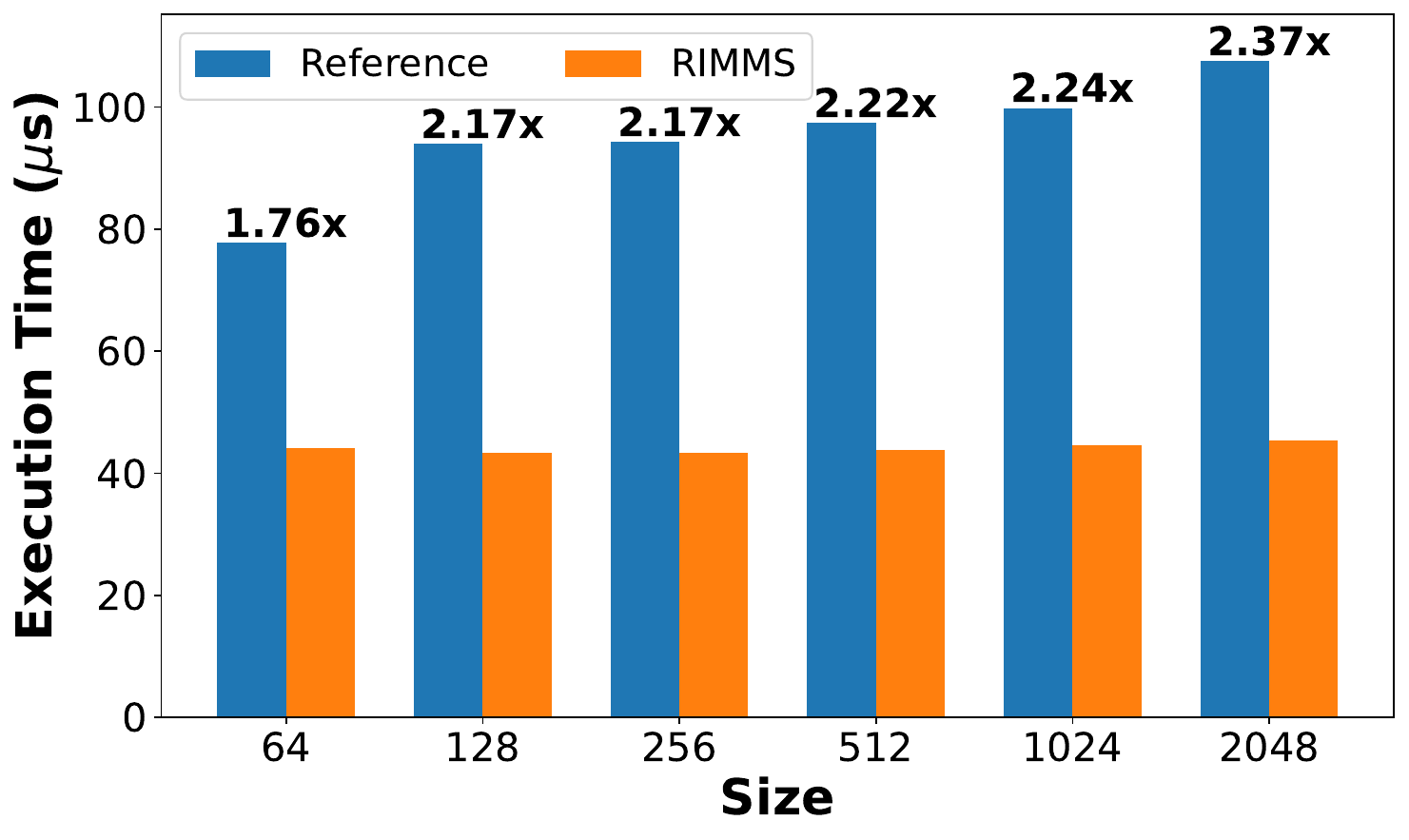}
        \caption{Both FFTs run on GPU}
    \end{subfigure}
    \caption{Execution time of 2FFT as a function of FFT size on Jetson AGX using the reference system and RIMMS.}
    \Description{To be updated after acceptance.}
    \label{fig:jetson-synthetic-2fft}
\end{figure}

\subsubsection{FPGA-Based Emulation}
We analyze execution time as a function of FFT size, ranging from 64 to 2,048 samples, for the 2FFT data flow, as illustrated in Figure~\ref{fig:zcu102-synthetic-2fft}. Our analysis compares the performance of RIMMS against the reference implementation across two scenarios, as shown in Figure~\ref{fig:zcu102-synthetic-2fft}. In the first scenario (CPU-ACC), FFT runs on the CPU, while IFFT runs on the accelerator (Figure~\ref{fig:zcu102-synthetic-2fft}(a)). In the second scenario (ACC-ACC), both FFTs are executed on the accelerator (Figure~\ref{fig:zcu102-synthetic-2fft}(b)). Each bar annotates the speedup achieved with RIMMS relative to the reference implementation. In our experiments, the data remains in its last computed location without additional memory transfers to the host CPU. 

In the CPU-ACC scenario, we consistently observe a speedup of approximately 1.3X across all sample sizes. In contrast, the speedup in the ACC-ACC scenario increases steadily from 2.07X to 4.66X as the sample size grows. In the CPU-ACC scenario, RIMMS reduces the total number of memory copies by one, while it eliminates three memory copies in the ACC-ACC scenario. We have excluded results for CPU-only execution since the execution times remain the same when the accelerator is not used. Similarly, in the ACC-CPU scenario, no change in execution time is observed because both the reference and RIMMS involve the same number of memory copies.

\subsubsection{GPU-Based SoC}
We compile and deploy the 2FFT synthetic application on the GPU-based SoC (Jetson AGX) to demonstrate the portability of the memory management functions and protocols. Similar execution time analysis is performed for FFT size across two scenarios: the CPU-GPU scenario, where the second FFT runs on the GPU, and the GPU-GPU scenario, where both FFT instances are executed consecutively on the GPU. In the CPU-GPU scenario, we observe a speedup of up to 1.96X, as shown in Figure~\ref{fig:jetson-synthetic-2fft}(a). 
In contrast, the speedup reaches up to 2.37X in the GPU-GPU scenario, as illustrated in Figure~\ref{fig:jetson-synthetic-2fft}(b). However, notable trends in the GPU system emerge that are absent in the FPGA-based system. For the CPU-GPU scenario, the speedup consistently decreases as the sample size grows, dropping from 1.96X to 1.09X. This reduction occurs because the CPU becomes a bottleneck as sample sizes increase, slowing down execution. This effect is more pronounced on the Jetson than the ZCU102 due to the different scales of execution times. In contrast, the speedup in the GPU-GPU scenario shows a slight increase as the sample size grows, rising from 1.76X to 2.37X.

\begin{table}[th]
    \centering
    \caption{2FZF execution time with respect to sample size in two deployment scenarios on ZCU102 FPGA and Jetson AGX. All times are given in $\mu$s scale, and the SpdUp column shows the speedup relative to the reference method. Bolded numbers show the first instance, where ACC-only execution becomes faster than CPU-only execution.}
    \renewcommand{\arraystretch}{1.2}
    \setlength{\tabcolsep}{2.4pt}
    \begin{tabular}{|c|c||c|c|c||c|c|c|}
    \hline
        \multirow{2}{*}{Size} & Exec     & \multicolumn{3}{c||}{ZCU102}                         & \multicolumn{3}{c|}{Jetson AGX}                     \\ \cline{3-8} 
                              & Type     & Reference          & RIMMS             & SpdUp       & Reference          & RIMMS             & SpdUp       \\ \hline \hline
        \multirow{2}{*}{32}   & CPU-only & 16.37              & \textbf{15.68}    & 1.04        & 4.02               & 3.98              & 1.01        \\ \cline{2-8}
                              & ACC-only & 19.25              & \textbf{10.79}    & 1.78        & 256.78             & 93.98             & 2.73        \\ \hline
        \multirow{2}{*}{64}   & CPU-only & 27.84              & 27.98             & 1.00        & 7.00               & 7.03              & 1.00        \\ \cline{2-8}
                              & ACC-only & 28.26              & 13.00             & 2.17        & 238.81             & 93.93             & 2.54        \\ \hline
        \multirow{2}{*}{128}  & CPU-only & \textbf{54.54}     & 54.45             & 1.00        & 13.81              & 13.79             & 1.00        \\ \cline{2-8}
                              & ACC-only & \textbf{46.65}     & 16.82             & 2.77        & 242.05             & 94.13             & 2.57        \\ \hline
        \multirow{2}{*}{256}  & CPU-only & 112.07             & 112.23            & 1.00        & 28.95              & 29.21             & 0.99        \\ \cline{2-8}
                              & ACC-only & 84.16              & 24.74             & 3.40        & 246.31             & 94.13             & 2.62        \\ \hline
        \multirow{2}{*}{512}  & CPU-only & 235.80             & 235.82            & 1.00        & 61.87              & 61.83             & 1.00        \\ \cline{2-8}
                              & ACC-only & 158.71             & 39.95             & 3.97        & 250.63             & 94.70             & 2.65        \\ \hline
        \multirow{2}{*}{1024} & CPU-only & 504.48             & 505.71            & 1.00        & 135.16             & \textbf{134.47}   & 1.01        \\ \cline{2-8}
                              & ACC-only & 307.44             & 71.04             & 4.33        & 257.36             & \textbf{95.44}    & 2.70        \\ \hline
        \multirow{2}{*}{2048} & CPU-only & 1,081.16           & 1,082.41          & 1.00        & \textbf{289.81}    & 289.13            & 1.00        \\ \cline{2-8}
                              & ACC-only & 604.78             & 132.13            & 4.58        & \textbf{267.98}    & 97.71             & 2.74        \\ \hline
    \end{tabular}
    \label{tab:2fzf}
\end{table}

\subsection{Experiments with the 2FZF Data Flow}

\begin{figure}[th]
    \centering
    \begin{subfigure}{0.75\textwidth}
        \centering
        \includegraphics[width=\textwidth]{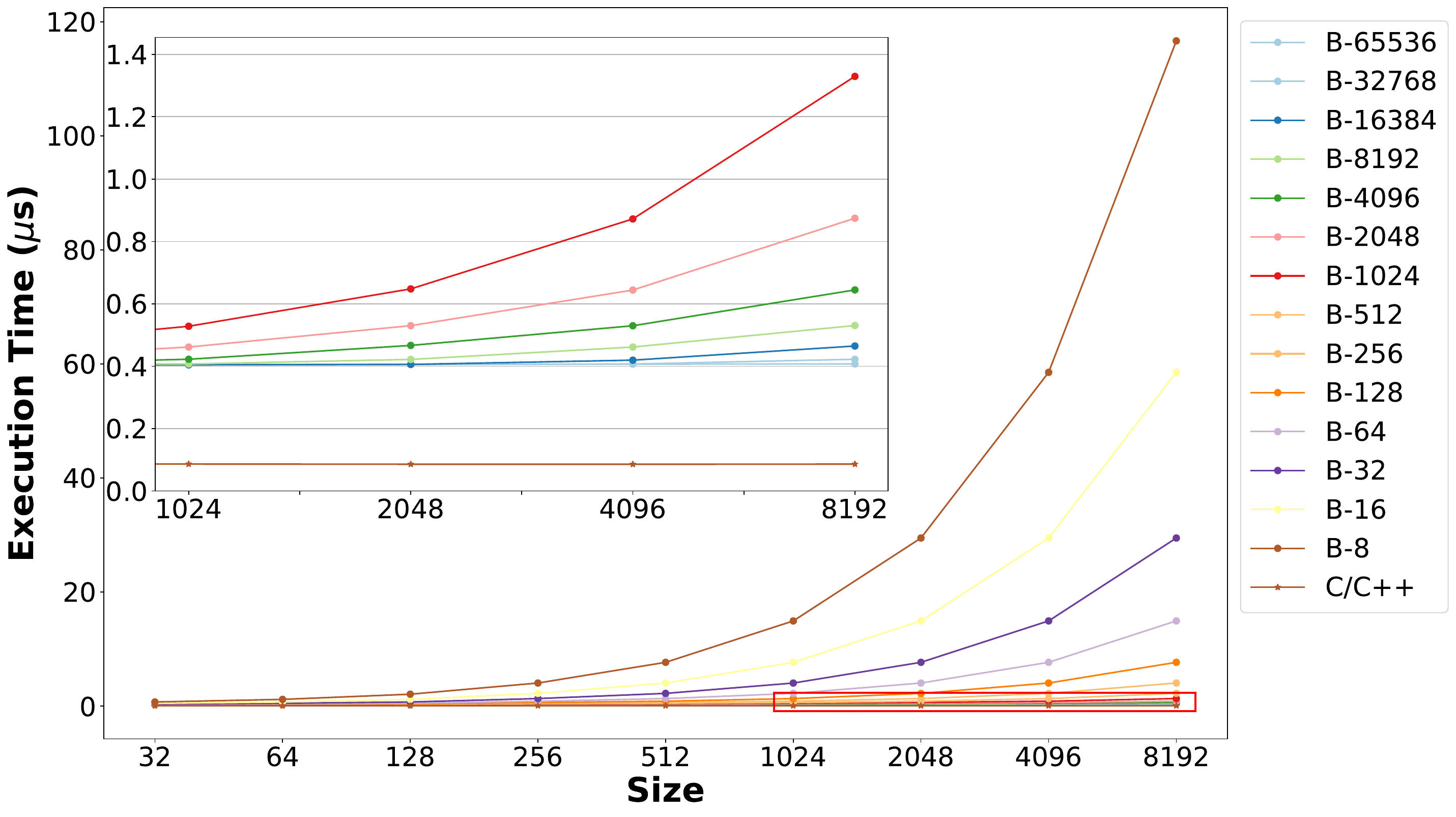}
        \caption{Time taken to allocate}
    \end{subfigure}
    \begin{subfigure}{0.75\textwidth}
        \centering
        \includegraphics[width=\textwidth]{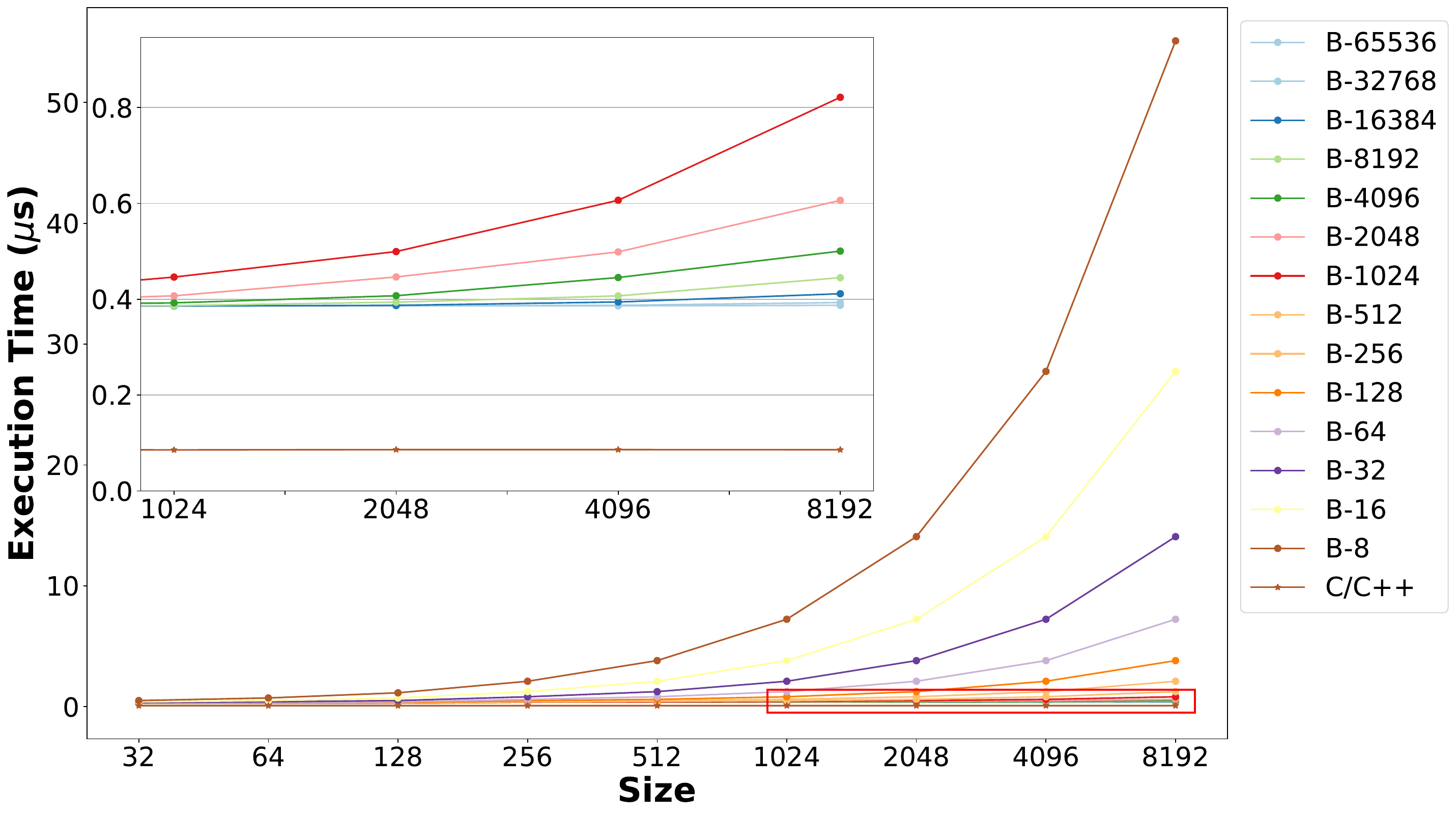}
        \caption{Time taken to deallocate}
    \end{subfigure}
    \caption{Memory management overhead across problem sizes for a given block size and comparison against the C/C++ default on (a) allocation and (b) deallocation. Inner plots focus on the regions sensitive to problem size and block size.}
    \Description{To be updated after acceptance.}
    \label{fig:allocation-deallocation}
\end{figure}

Next, we focus on the 2FZF, which has a more complex data flow with two accelerator types (FFT and ZIP). To isolate the effects of memory management optimizations from other performance factors, we execute the first two FFTs in 2FZF sequentially, eliminating the benefits of parallel execution. We analyze its execution time when the application is implemented with hardware-agnostic memory management functions and is compiled and executed using the proposed memory management protocols. Table~\ref{tab:2fzf} shows the execution times for both CPU-only and accelerator (ACC)-only versions on the ZCU102 and Jetson platforms. In these experiments, the FFT and ZIP kernels are executed using dedicated accelerators on the ZCU102 board, and using GPU on the Jetson platform. For the CPU-only configuration, there is a negligible difference in execution time when compared to the reference implementation on both FPGA and GPU platforms. This is an important outcome as it confirms that the RIMMS protocols that are integrated with the runtime do not introduce any overhead. However, in the ACC-only execution, we see performance improvements of up to 4.58X on the ZCU102 and up to 2.74X on the Jetson platform. The speedup trends on both platforms are consistent with those observed in the 2FFT experiment. On the ZCU102 platform, the ACC-only execution is slower than CPU-only execution when the sample size is 32 but performs better with sample sizes of 128 or larger, achieving up to 1.79X speedup compared to CPU-only with the sample size of 2,048. With RIMMS, the ACC-only version is faster than the CPU-only execution, even at a sample size of 32. The speedup grows to 8.2X for a sample size of 2,048. On the GPU platform, the reference implementation benefits from GPU acceleration, but a speedup is observed only when the problem size reaches 2,048, likely due to the overhead of data transfer between CPU and GPU. This overhead is mitigated with RIMMS, and speedup occurs earlier, starting at a sample size of 1,024.

\subsubsection{Allocation/Deallocation Overhead Analysis}\label{sec:results_bitset_overhead}
The proposed hardware-agnostic memory management functions enable application developers to use any available PEs on the target system without binding the data with specific hardware. However, this flexibility comes with an overhead as protocols described in Section~\ref{subsec:mmprotocols}
involve processes to monitor and identify the data owner. We analyze the overhead associated with the hardware agnostic memory allocation and deallocation functions \cedrmalloc~and \cedrfree~using Figure~\ref{fig:allocation-deallocation}(a) and Figure~\ref{fig:allocation-deallocation}(b), respectively. These functions operate at the block level as described in Section~\ref{subsec:mmprotocols}. Smaller block sizes offer the ability to match the granularity of the requested data size, achieve better memory utilization, and meet the needs of more allocation requests. However, a smaller block size results in more blocks to search and manage, increasing the memory management overhead. Larger block sizes reduce the search space and result in faster memory management with a trade-off in memory utilization efficiency. For the analysis, we fix the block size and plot the time taken to allocate or deallocate arrays of floating-point variables, with sizes ranging from 32 to 8,192 elements. We vary the block sizes from 8 bytes to 65,536 bytes, and trend lines for each block size are overlaid in the corresponding plots. Referring to Figure~\ref{fig:allocation-deallocation}, when problem size is small (32 elements), the overhead is not sensitive to the block size. As the problem size grows, the overhead of smaller block sizes increase rapidly. Focusing on the zoomed in portions of the two plots, for the problem size of 8,192, block size of 4,096 offers a compromise with \cedrmalloc~and \cedrfree~taking 0.64$\mu$s and 0.5$\mu$s while standard \emph{malloc} and \emph{free} functions take 0.86$\mu$s and 0.86$\mu$s, respectively.  

\subsubsection{Runtime Overhead Analysis}
We utilize a microbenchmark which performs the \emph{last resource flag} check iteratively one million times. We measure the runtime cost introduced by RIMMS based on the per-call overhead of the APIs. This overhead arises solely from a simple flag-checking operation integrated into each API call for the \emph{last resource flag} of each input. The check involves a single table lookup followed by a conditional branch to determine whether memory copies are required. Our measurements on the ZCU102 platform with 1 million API inputs show that this flag-checking overhead takes an average of \textit{1.16 CPU cycles per input}, with a range of 1 to 2 cycles. Given the low cost and the fact that this check occurs only at API boundaries per input, its impact on end-to-end application execution time is \textit{negligible}. This minimal overhead enables flexible memory management by eliminating memory copies without sacrificing application performance.

\subsection{Experiments with the 3ZIP Data Flow}
\begin{figure}[t]
    \centering
    \includegraphics[width=\textwidth]{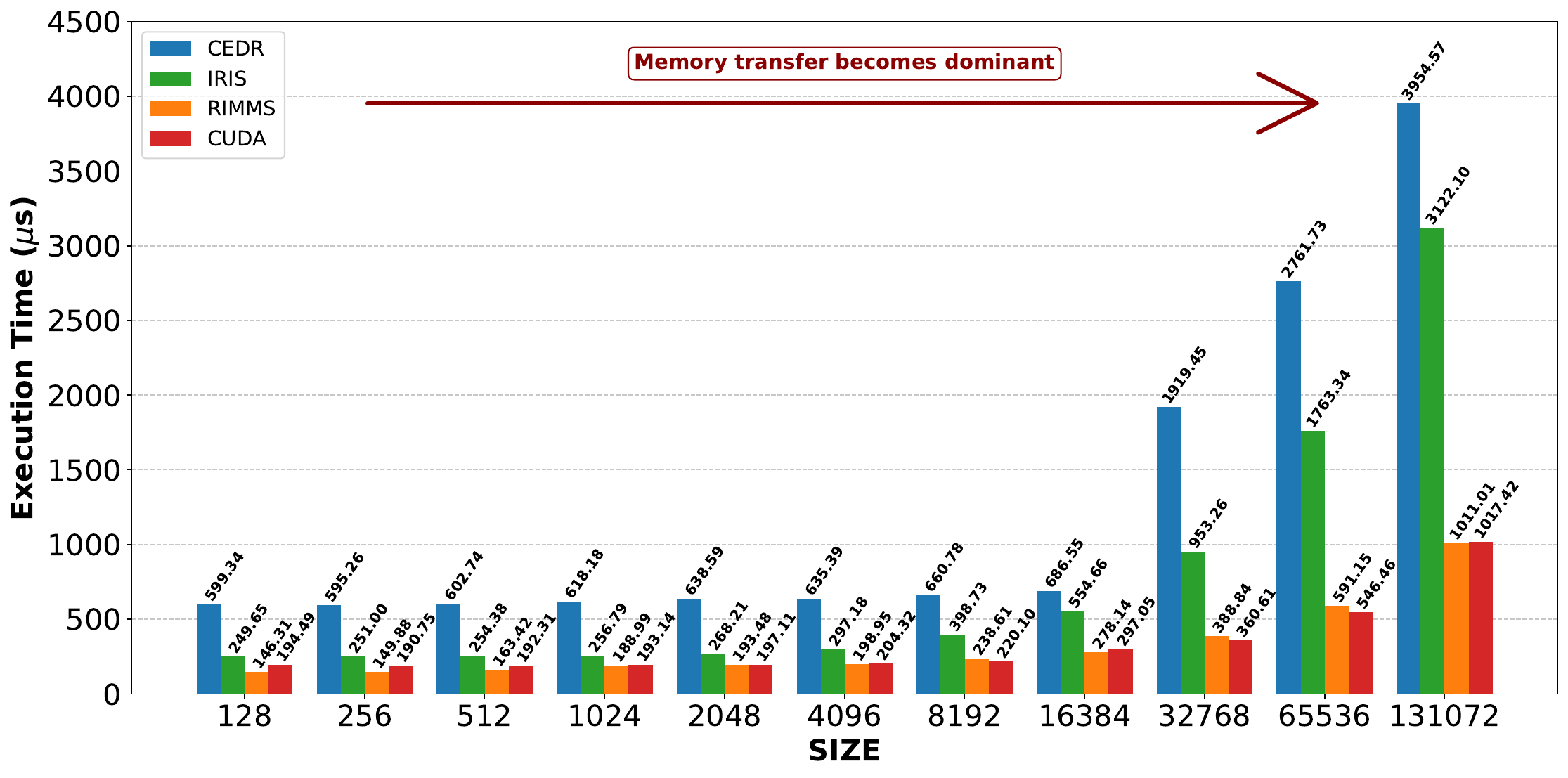}
    \caption{Execution time of 3ZIP using CEDR, IRIS, RIMMS, and native CUDA as a function of ZIP size on Jetson AGX platform.}
    \Description{To be updated after acceptance.}
    \label{fig:3ZIP}
\end{figure}
Until this point, RIMMS has been evaluated primarily against CEDR, which we have used as a baseline. In this section, we extend the comparison to include other frameworks, specifically IRIS~\cite{runtime-TPDS-IRIS24} and native CUDA~\cite{CUDA_2008}, on the Jetson AGX platform using the 3ZIP application, as shown in Figure~\ref{fig:3ZIP}. We select 3ZIP because it can be scaled to arbitrary problem sizes, includes inter-kernel dataflow dependencies, and can be implemented consistently across all frameworks, enabling a fair evaluation. All frameworks (CEDR, IRIS, and RIMMS) are configured to use only the GPU as the computational resource to maintain consistency with the CUDA implementation. The CUDA version serves as an optimized comparison point and is designed to avoid transferring intermediate data between ZIP stages back to host memory, mirroring the approach taken by RIMMS. We sweep ZIP input sizes from $2^7$ to $2^{17}$. A detailed timing analysis of the CUDA implementation shows that as the problem size increases, the application becomes increasingly memory-bound. Specifically, the kernel-to-memory operation ratio increases from approximately 1:2 at $2^7$ elements to around 1:5 at $2^{17}$ elements. These ratios exclude the internal four memory copies required between ZIPs, meaning frameworks that perform these redundant copies would exhibit even more noticeable memory bottlenecks. RIMMS demonstrates consistent performance advantages compared to CEDR, achieving speedups ranging from 2.46X to 4.93X, primarily due to its elimination of redundant memory transfers and its ability to dynamically track memory usage across stages. When compared to IRIS, RIMMS delivers performance improvements ranging from 1.35X to 3.08X, demonstrating its efficiency in comparison to a modern, heterogeneous-capable runtime. Most notably, RIMMS closely tracks the performance of hand-crafted native CUDA implementation across all input sizes, with negligible differences, despite operating at a much higher level of abstraction. Developers using RIMMS benefit from a simplified, hardware-agnostic API that eliminates the need for explicit memory allocation, manual data transfer management, or writing device-specific code. As a result, RIMMS offers a substantial productivity advantage over both IRIS and CUDA, enabling portable, high-performance application development without compromising efficiency.

\subsection{Experiments with Real-world Applications}
Finally, we validate RIMMS and assess its performance by executing three real-world signal processing applications on the Jetson AGX platform, as shown in Table~\ref{tab:apps}. The RC application follows a task-level data flow identical to 2FZF (Figure~\ref{fig:synthetic-application}(b)) with a sample size of 256. The PD application has a similar data flow but scales computations to 128 parallel instances of 2FZF for a sample size of 128. The SAR application involves two consecutive FZF data flow phases, similar to 2FZF, with a sample size of 256, with 512-way parallelism in the first phase, and 256-way parallelism, with a sample size of 512 in the second phase. Each application includes pre- and post-processing functions around API calls or execution phases that are unsuitable for accelerator-based execution and are, therefore, run on the CPU. We profile the execution time of each configuration, where applications are implemented using the proposed memory management functions and deployed through the RIMMS. We compare the performance to the reference implementation using two hardware configurations: GPU-only and a 3-CPU, 1-GPU setup. The GPU-only results (Table~\ref{tab:apps}) validate the data presented in Table~\ref{tab:2fzf} for matching sample sizes (128 and 256). We use the round-robin scheduler to control task assignments. On the 3-CPU, 1-GPU setup, the N-way parallel tasks are scheduled in batches of four: the first three to individual CPU cores, and the fourth to the GPU in each round. This configuration allows us to validate the performance improvements gained through RIMMS, while correlating with the data in Table~\ref{tab:2fzf}. 

\begin{table}[t]
    \centering
    \caption{Evaluation of RIMMS with three signal processing applications on Jetson AGX platform.}
    \renewcommand{\arraystretch}{1.2}
    \setlength{\tabcolsep}{3pt}
    \begin{tabular}{|c|c|c|c|c|}
    \hline
        Application           & Configuration & Reference ($ms$) & RIMMS ($ms$) & Speedup \\ \hline \hline
        \multirow{2}{*}{RC}   & GPU-Only      & 1.53             & 1.32         & 1.16        \\ \cline{2-5}
                              & 3CPU-1GPU     & 1.19             & 1.23         & 0.97        \\ \hline
        \multirow{2}{*}{PD}   & GPU-Only      & 135.36           & 69.41        & 1.95        \\ \cline{2-5}
                              & 3CPU-1GPU     & 38.38            & 27.79        & 1.38        \\ \hline
        \multirow{2}{*}{SAR}  & GPU-Only      & 573.24           & 235.78       & 2.43        \\ \cline{2-5}
                              & 3CPU-1GPU     & 179.14           & 167.13       & 1.07        \\ \hline
    \end{tabular}
    \label{tab:apps}
\end{table}

\noindent \textbf{GPU-Only Setup:} With RIMMS, the RC application achieves a 1.16X speedup over the reference implementation on the GPU-only configuration, while the 2FZF achieves a 2.62X speedup. The lower speedup for RC is due to the increased proportion of non-API regions in the total execution time for this short-latency task, leading to smaller gains from memory management optimizations. As the computational complexity of the applications increases, the time spent on serial tasks decreases, resulting in improved speedups: 1.95X for the PD and 2.43X for the SAR, which aligns with the 2FZF's 2.57X improvement for sample sizes of 128, 256, and 512.

\noindent \textbf{3 CPUs-1 GPU Setup:} We observe no significant execution time difference for RC between the reference and RIMMS setups. Since RC consists of four tasks, with only the final task offloaded to the GPU, the total number of memory transfers remains unchanged. The PD application achieves a 1.38X speedup, as 75\% of the 128 2FZF flows are executed on the CPU, while only 25\% utilize the GPU due to the round-robin scheduling. This limits the speedup to around 1.38X, roughly a quarter of the speedup observed for 2FZF in Table~\ref{tab:2fzf} with a sample size of 128. In contrast, for the SAR application, performance gains from memory management are outweighed by those from parallel execution, reducing the overall speedup.

These findings validate that the trends observed in reference data flow applications also hold in real-world applications. Additionally, transitioning from GPU-only to 3CPU-1GPU setup decreases the overall execution time across all applications, as expected for the reference and RIMMS-based deployments. 

\subsection{Experiments with NF-based Approach on ZCU102}

\begin{figure}[t]
    \centering
    \includegraphics[width=\textwidth]{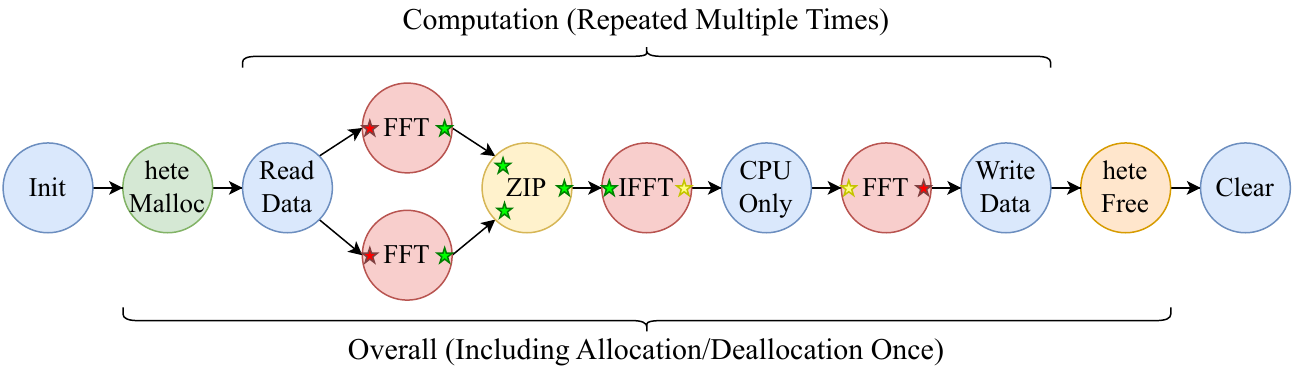}
    \caption{PD application DAG showing possible memory locations before and after FFT and ZIP nodes. Red stars show the initial API entrances and the last API exit where data has to be moved. Yellow stars show the entrance and exit of a CPU-Only region where data has to be moved to and from the CPU memory. Green stars show the locations where RIMMS can eliminate redundant memory copies.}
    \Description{To be updated after acceptance.}
    \label{fig:PD_DAG}
\end{figure}

\subsubsection{Real Application}
In this experiment, we focus on a specific application, PD, since it is easy to represent as a DAG and complex enough to cover multiple different types of execution. 
Many real-world applications allocate memory once and reuse it to process multiple inputs, and PD follows this pattern.
Based on this behavior, we categorize the PD application into distinct regions, as illustrated in Figure~\ref{fig:PD_DAG}. The \emph{Init} and \emph{Clear} nodes handle initialization and cleanup operations, excluding memory allocation and deallocation. The \emph{Computation} region represents the core processing flow of a single input, covering its acquisition, transformation, and final output relay. This region excludes performing any memory allocation or deallocation. The \emph{Overall} region encompasses both memory allocation and deallocation in addition to the execution of the \emph{Computation} region. Since the \emph{Computation} region is responsible for processing each input independently, it is repeated for every input. If an application processes $N$ inputs, the \emph{Computation} region executes $N$ times, while the \emph{Overall} region spans the execution of all $N$ \emph{Computation} instances along with the associated memory allocation and deallocation steps. \emph{Init} and \emph{Clear} are excluded from the \emph{Overall} region, as they involve OS-related I/O operations such as file reads and writes, which introduce fluctuations in execution time and lead to inconsistent results. All results presented in this section are obtained using the ACC-Only configuration on the ZCU102 platform.

\subsubsection{Allocation Overhead for PD}\label{subsec:allocation_overhead_PD}

\begin{figure}[t]
    \centering
    \includegraphics[width=\textwidth]{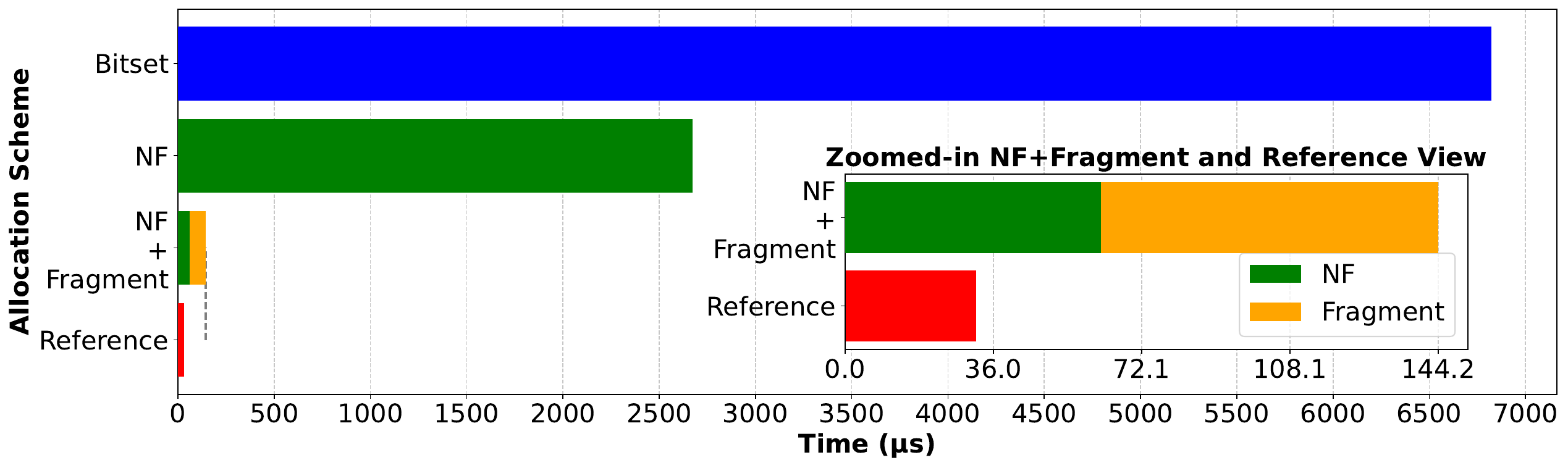}
    \caption{Allocation overheads when using different schemes with the PD application on ZCU102 platform.}
    \Description{To be updated after acceptance.}
    \label{fig:allocation_overheads}
\end{figure}

Examining the \emph{Computation} region of the PD application more closely, we identify eight distinct data points required for the processing flow (corresponding to the number of edges in Figure~\ref{fig:PD_DAG}). Figure~\ref{fig:allocation_overheads} presents the overhead of different marking systems implemented as part of RIMMS for the PD application. We use a block size of 4,096 for the bitset-based allocation, as mentioned in Section~\ref{sec:results_bitset_overhead}. Compared to the bitset-based approach, the NF-based allocation reduces the overhead by 2.55X. Since each FFT and ZIP node consists of 128 parallel nodes, both approaches require 128 separate \cedrmalloc~calls per data point. However, when the \emph{fragment} function is utilized with the NF-based allocation, a single \cedrmalloc~call is enough, reducing the number of allocation and deallocation calls to one per data point. As a result, we observe an overhead reduction of 18.53X compared to the NF-only implementation, where \cedrmalloc~and \emph{fragment} call contribute to the overhead by 43.07\% and 56.93\% respectively. Coupling NF with \emph{fragment} call reduces the timescale of the overhead from millisecond to microsecond, and results with a negligible overhead compared to the reference implementation. Using this coupled approach, the benefit of RIMMS in eliminating redundant memory copies becomes more evident in real-world scenarios, as presented in the following subsection.

\begin{table}[t]
    \centering
    \caption{Computation only and overall execution times (in $ms$) of PD application using reference method and RIMMS on ZCU102 platform. SpeedUp (SpdUp) achieved by using RIMMS relative to the reference method. SpdUp cells in the Overall part are colored based on the difference of the SpdUp of Overall results relative to the Computation Only results. Green: $=0$; Orange: $\leq0.02$; Yellow: $\leq0.05$; Red: $>0.05$.}
    \renewcommand{\arraystretch}{1.2}
    \setlength{\tabcolsep}{1.5pt}
    \begin{tabular}{|c||c|c|c||c|c|c|c|c|c|c|}
    \hline
        \multirow{3}{*}{Repeat} & \multicolumn{3}{c||}{Computation Only}                                          & \multicolumn{7}{c|}{Overall}                                                                                                                                       \\ \cline{2-11}
        \multirow{3}{*}{Count}  & \multirow{3}{*}{Reference} & \multirow{3}{*}{RIMMS} & \multirow{3}{*}{SpdUp}   & \multirow{3}{*}{Reference} & \multicolumn{6}{c|}{RIMMS}                                                                                                             \\ \cline{6-11}
                                &                            &                        &                          &                            & \multirow{2}{*}{Bitset} & \multirow{2}{*}{SpdUp} & \multirow{2}{*}{NF} & \multirow{2}{*}{SpdUp} & NF $+$      & \multirow{2}{*}{SpdUp} \\
                                &                            &                        &                          &                            &                         &                        &                     &                        & Fragment    &                        \\ \hline\hline
        1     & 6.74     & 4.03     & 1.67 & 7.02     & 11.28    & \cellcolor[HTML]{FFCCCB}0.62 & 6.93     & \cellcolor[HTML]{FFCCCB}1.01 & 4.34     & \cellcolor[HTML]{FFFF00}1.62 \\\hline
        10    & 61.60    & 33.90    & 1.82 & 61.90    & 41.15    & \cellcolor[HTML]{FFCCCB}1.50 & 36.80    & \cellcolor[HTML]{FFCCCB}1.68 & 34.24    & \cellcolor[HTML]{FFD580}1.81 \\\hline
        50    & 299.91   & 164.47   & 1.82 & 300.21   & 171.72   & \cellcolor[HTML]{FFCCCB}1.74 & 167.37   & \cellcolor[HTML]{FFFF00}1.79 & 164.82   & \cellcolor[HTML]{90EE90}1.82 \\\hline
        100   & 600.71   & 333.25   & 1.80 & 601.02   & 340.50   & \cellcolor[HTML]{FFFF00}1.76 & 336.15   & \cellcolor[HTML]{FFD580}1.78 & 333.60   & \cellcolor[HTML]{90EE90}1.80 \\\hline
        1,000 & 5,955.07 & 3,267.73 & 1.82 & 5,955.39 & 3,276.98 & \cellcolor[HTML]{FFD580}1.81 & 3,272.63 & \cellcolor[HTML]{FFD580}1.81 & 3,270.09 & \cellcolor[HTML]{90EE90}1.82 \\\hline
    \end{tabular}
    \label{tab:PD_ZCU102_repeated_exec}
\end{table}

\subsubsection{Overall Application Profile}

Looking at the \emph{Computation Only} results in Table~\ref{tab:PD_ZCU102_repeated_exec}, we observe that using RIMMS and eliminating redundant memory copies results in a speedup of up to 1.82X. In a real-world scenario, memory allocation and deallocation also contribute to execution time. Ideally, with enough repetitions in the computational region, the impact of allocation overhead should diminish, causing the speedup achieved from the Overall execution approach to the \emph{Computation Only} speedup. To evaluate this, we compare the \emph{Overall} execution times of the reference method and the three different RIMMS allocation schemes. Starting with the bitset-based allocation, we observe variations in speedups for up to 100 repeats, including an initial slowdown for a single execution. However, beyond 100 repetitions, the speedup reaches closer to the \emph{Computation Only} results. Next, for the NF-based allocation, there is no initial slowdown, even with a single execution. However, the achieved speedup remains below the \emph{Computation Only} results until about 50 repeats. Even with 1,000 repetitions, this method does not fully match the expected speedup. Finally, with the NF-based allocation combined with the \emph{fragment} call, this approach's lower overhead allows for speedup values closer to the \emph{Computation Only} results from the very first execution. From 50 repeats onward, this method achieves the exact speedup seen in the \emph{Computation Only} results.

\section{Conclusion}\label{sec:conclusion}
As systems converge to higher degree of heterogeneity, the complexity of managing hardware imposes a significant burden on runtime designers. As a result, there is a growing demand on compiler designers to embed comprehensive information such as data flow, dependency analysis, and hardware-specific representations of application tasks into their binaries. In this study, we present RIMMS, a runtime memory management system that enables data flow aware application deployment on heterogeneous systems without requiring users to have expertise on the memory hierarchy of the target system while developing their applications. We develop portable and hardware-agnostic memory management primitives that allow the runtime system to track the location of the data such that the scheduler has the flexibility to make task-to-PE mapping decisions based on the state of the system resources and the location of the data. To further improve computational efficiency in systems with FPGAs, we introduced an NF-based marking system, which reduces allocation overhead compared to the bitset-based method while maintaining flexibility. Additionally, the introduction of a \emph{fragment} function allows structured memory reuse, eliminating unnecessary allocations and reducing the impact of allocation overhead in real applications. Our results demonstrate that this combination achieves near-optimal speedups, up to 1.82X, by accelerating allocation and reducing redundant memory copies. Such advancements enable closing the loop on the design of heterogeneous compilers, runtimes, and hardware. As future work, our aim is to develop contention-aware intelligent algorithms for memory allocation on heterogeneous systems. This will enable a wide range of performance gains within and across accelerator boundaries.

\begin{acks}
This material is based on research sponsored by Air Force Research Laboratory (AFRL) and Defense Advanced Research Projects Agency (DARPA) under agreement number FA8650-18-2-7860. The U.S. Government is authorized to reproduce and distribute reprints for Governmental purposes notwithstanding any copyright notation thereon. The views and conclusions contained herein are those of the authors and should not be interpreted as necessarily representing the official policies or endorsements, either expressed or implied, of AFRL and DARPA or the U.S. Government. 

We appreciate the continuous and generous support from the AMD University Program, including the donation of FPGA prototyping board used in this work.

Dr. Akoglu and Dr. Ogras have disclosed an outside interest in DASH Tech IC to the University of Arizona and University of Wisconsin, respectively.  Conflicts of interest resulting from this interest are being managed by the respective universities in accordance with their policies.
\end{acks}

\bibliographystyle{ACM-Reference-Format}
\bibliography{refs}

\end{document}